\documentclass[sigconf]{acmart}

\usepackage{graphicx,dsfont,amssymb,booktabs} 

\usepackage[noend]{algpseudocode}
\usepackage{algorithm}
\usepackage{amsmath}
\usepackage{diagbox}
\usepackage{subcaption} 

\usepackage{color, colortbl}
\usepackage{tabulary}
\usepackage{tabularx}
\usepackage{makecell}
\usepackage[textsize=scriptsize]{todonotes}
\usepackage{multirow}
\usepackage{ctable} 

\usepackage{pgfplots}
\usepackage{tikz}
\usetikzlibrary{
	arrows.meta,
	colorbrewer,
	decorations.pathreplacing,
	automata, 
	positioning }

\usepackage{listings}
\usepackage{balance}

\AtBeginDocument{%
  \providecommand\BibTeX{{%
    \normalfont B\kern-0.5em{\scshape i\kern-0.25em b}\kern-0.8em\TeX}}}

\acmYear{2020}\copyrightyear{2020}
\setcopyright{acmcopyright}
\acmConference[DEBS '20]{The 14th ACM International Conference on Distributed and Event-based Systems}{July 13--17, 2020}{Virtual Event, QC, Canada}
\acmBooktitle{The 14th ACM International Conference on Distributed and Event-based Systems (DEBS '20), July 13--17, 2020, Virtual Event, QC, Canada}
\acmPrice{15.00}
\acmDOI{10.1145/3401025.3401742}
\acmISBN{978-1-4503-8028-7/20/07}

\newcommand{\framework}{hSPICE}

\newcommand\bbGammaVar{\reflectbox{\rotatebox[origin=c]{180}{$\mathbb L$}}}

\usepackage[most]{tcolorbox}
\fancypagestyle{firstpage}{%
	\rhead{}
	\lhead{}
	\cfoot{	
		\begin{tcolorbox}[colback=white,
			colframe=red,
			width=13cm,
			arc=3mm, auto outer arc,
			]
			{\color{red}
				\copyright ~ACM, [2020]. This is the author's version of the work. It is posted here by permission of ACM for your personal use. Not for redistribution. The definitive version  will be published in proceedings of the 14th ACM International Conference on Distributed and Event-based Systems (DEBS '20), July 13--17, 2020, Virtual Event, QC, Canada.}
		\end{tcolorbox}
	}
}

\begin{document}

\title{\framework: State-Aware Event Shedding in Complex Event Processing}
\renewcommand{\shorttitle}{\framework}


\author{Ahmad Slo, Sukanya Bhowmik,  Kurt Rothermel}
\orcid{1234-5678-9012}
\affiliation{%
	\institution{University of Stuttgart}
}
\email{firstName.lastName@ipvs.uni-stuttgart.de}



\begin{abstract}
 In complex event processing (CEP), load shedding is performed to maintain a given latency bound during overload situations when there is a limitation on resources. However, shedding load implies degradation in the quality of results (QoR). Therefore, it is crucial to perform load shedding in a way that has the lowest impact on QoR. Researchers, in the CEP domain, propose to drop either events or partial matches (PMs) in overload cases. They assign utilities to events or PMs by considering 
 either the importance of events or the importance of PMs but not both together.
 In this paper, we propose a load shedding approach for CEP systems  that combines these approaches by assigning a utility to an event by considering both the event importance and the importance of PMs. We adopt a probabilistic model that uses the type and position of an event in a window and the state of a PM to assign a utility to an event corresponding to each PM. 
 We, also, propose an approach to predict a utility threshold that is used to drop the required amount of events to maintain a given latency bound. By extensive evaluations on two real-world datasets and several representative queries, we show that, in the majority of cases, our load shedding approach outperforms state-of-the-art load shedding approaches, w.r.t. QoR.
\end{abstract}



\maketitle
\thispagestyle{firstpage}

\section{Introduction}
Complex event processing (CEP) systems are used in many applications  to detect patterns in input event streams  \cite{spectre:2017, Balkesen:2013:RRI:2488222.2488257, Zacheilas:2015}.
The criticality of  detected patterns (also called complex events) depends on the application. For example, in fraud detection systems in banks, detected complex events might indicate that a fraudster tries to withdraw money from a victim's account. Naturally, the complex events in this application are critical. On the other hand, in applications like network monitoring, soccer analysis, and transportation \cite{Olston:2003:AFC:872757.872825, espice, pspice}, the detected complex events might be less critical. As a result, these applications might tolerate imprecise detection or loss of some complex events.

In CEP systems, input events are streamed continuously to CEP operators where the input events (or simply events) are partitioned into windows of events. Events within windows are processed by CEP operators  to detect patterns (called pattern matching). For most applications, it is important to detect complex events within a certain latency bound (LB) where the late detected complex events become useless \cite{Quoc:2017:SAC:3135974.3135989, 10.1145/3361525.3361551}. However, if the rate of input events exceeds the processing capacity of CEP operators, the input events queue up and the  detection latency of complex events increases, possibly resulting in violation of the given latency bound. For CEP applications that tolerate imprecise detection of complex events and have limited processing resources, one way to keep the given latency bound is by using load shedding \cite{espice, pspice, He2014OnLS, bo:2020}. Load shedding reduces the overload on a CEP operator by either dropping events from the operator's input event stream or by dropping a portion of the operator's internal state. This results in decreasing the number of queued events and in increasing the operator processing rate, hence maintaining the given latency bound.    

Of course, load shedding may impact the quality of results (QoR) as it might falsely drop complex events (denoted by false negatives) or/and falsely detect complex events (denoted by false positives). Therefore, it is crucial to shed load with minimum adverse impact on QoR.
In \cite{espice, He2014OnLS}, the authors propose two \textit{black-box} load shedding approaches for CEP systems where their approaches drop input events that have the lowest utility. The approach in \cite{espice} uses event type and position within windows as features to probabilistically learn about the utility of events in windows.  In \cite{He2014OnLS}, the event utility depends on the frequency of events in patterns and in the input event stream. In \cite{pspice, bo:2020}, the authors propose two \textit{white-box} approaches to perform load shedding in CEP where the focus is on dropping partial matches. A partial match is a detected part of a pattern that could become a complex event if the full pattern is matched. However, the approach in \cite{bo:2020} might also drop input events if the given latency bound might be violated. Both approaches depend on the following features to learn about the utility of PMs: the progress/state of the PM in the window and the number of remaining events in the window.  These two features are used to predict the completion probability and the processing cost of the PMs and hence the PM utilities.

In the \textit{black-box} approach, load shedding is performed  in a finer granularity (event granularity), i.e., it  drops individual events from windows, in comparison to \textit{white-box} dropping approaches which mainly drop PMs, i.e, dropping in a coarser granularity. As a result, the white-box approaches might drop PMs that have relatively high utilities which adversely impacts QoR  even if there exist events that may be dropped without impacting QoR.
On the other hand, the black-box approaches neither consider the importance nor the state of PMs. An event might have different utilities for individual PMs, depending on the importance and the state of PMs.
Thus, in this paper, we propose a new white-box load shedding strategy called \framework~ that combines the best of both black-box and white-box approaches.

In particular, \framework~ is  a white-box load shedding approach that drops events from PMs-- it sheds on the event-granularity-- while considering the operator's internal  state.  
\framework~  predicts the utility of the events  using a probabilistic model. The model uses the event type, the event position within a window, and the state of partial matches in a window to learn about the utility of events within windows.
An important factor that influences the effectiveness of a load shedding approach is its overhead in performing the load shedding.
A high load shedding overhead implies that a high percentage of the available processing power will be used to take the shedding decision. This results in reducing the available processing power to perform pattern matching, thus adversely impacting QoR.
As we will show, \framework~ is a lightweight, efficient load shedding approach.

More specifically, our contributions in this paper are as follows:
\begin{itemize}
	\item 
	We propose a white-box load shedding approach for complex event processing called \framework. \framework~ performs load shedding by dropping events from PMs.
	\framework~ uses a probabilistic model to learn the utility of an event \textit{for a PM} within a window. As learning features, we use the type and position of the event within the window and the state of the PM.

	\item
	We provide an algorithm to estimate the number of events to drop to maintain the given latency bound. Additionally, we propose an approach that enables \framework~ to perform load shedding in a lightweight manner.   
		
	\item 
	We provide extensive evaluations on two real-world datasets  and a representative set of CEP queries to prove the effectiveness of \framework~ and to show its performance, w.r.t. its adverse impact on QoR,   in comparison to state-of-the-art load shedding approaches.
		
\end{itemize}

\section{Preliminaries and Problem Statement}
\label{sec:background}

\subsection{Complex Event Processing}
A CEP system consists of a set of operators that are connected in the form of a directed acyclic graph (DAG). An operator in a CEP system correlates input events to detect patterns. The detected patterns are called complex events. An event in the input event stream (denoted by $S_{in}$) consists of  meta-data and attribute-value pairs. The meta-data contains event type, sequence number and/or timestamp, while the attribute-value pairs represent the event data.  For example, the type (denoted by $T_e$) of event $e$ might represent a company name in a stock application, a player ID in a soccer application, or a bus ID in a transportation application. The event data might contain stock quotes, player positions, or bus locations in these applications. Events in the input event streams have global order, for example, by using the sequence number or the timestamp and a tie-breaker.

Our focus in this paper is on CEP systems consisting of a single operator, where the operator matches one or more patterns (i.e., multi-query). We define the set of patterns that the operator matches as $\mathbb{Q}= \{q_i : 1 \leq i \leq n\}$, where $n$ is the number of patterns. Since patterns might have different importances, each pattern has a weight reflecting its importance. The patterns' weights are determined by a domain expert and they are defined as follows: $\mathbb{W_Q}= \{ w_{q_i} : 1 \leq i \leq n\}$, where $w_{q_i}$ is the weight of pattern $q_i$.  
In CEP systems, the input event stream $S_{in}$ is continuous and infinite, where the input event stream is partitioned into windows of events. Windows in CEP are opened depending on predicates such as time-based, count-based, or logical predicates. Moreover, the length of windows might be defined by time, event count, or logical conditions \cite{Balkesen:2013:RRI:2488222.2488257, Tatbul:2006:WLS:1182635.1164196}. The number of events in a window is defined as window size (denoted by $ws$). 
Each event in window $w$ has a position where the position $P_e$ of event $e$ represents the number of events that precedes event $e$ in window $w$.   
Windows might overlap which means that there may exist  more than one open window at the same time. Hence, event $e$ might belong to multiple windows, where it has different positions $P_e$ within different windows.
To clarify the system model, let us introduce the following example. 

\begin{figure*}[t]
	\centering
	\resizebox{0.99\linewidth}{!}{

\definecolor{w1Color}{rgb}{1,0,0}
\definecolor{w2Color}{rgb}{137,0,255}
\definecolor{w3Color}{rgb}{0,0,1}
\tikzset{
	arrow style right/.style={color=black, line width=1pt, -{>[scale=2.5, length=4, width=2]}}, 
	arrow style left/.style= {color=black, line width=1pt, {<[scale=2.5, length=4, width=2]}-}, 
	node pattern style/.style= {shorten >=1pt, node distance=3cm,on grid,auto, line width=1pt,  initial distance= 0.5cm},	
	node pm style/.style= { shorten >=1pt, node distance=2 cm,on grid,auto, line width=1pt},	
	big font/.style={font=\fontsize{8.0}{9.0}\selectfont},
	every node/.style={big font},
	every state/.style={minimum size=0pt, inner sep=2pt}
}

\newcommand{\xStart}{-18}
\newcommand{\xEnd}{-9}
\newcommand{\yStart}{-5}

\newcommand{\xEventShift}{-0.8}

\begin{tikzpicture}[xscale= 1,yscale=0.6]

\begin{scope}[node pattern style]
	\node[ state,initial]  (s_0) at (\xStart, - 1)  {$s_0$}; 
	\node[ state] (s_1) [right=of s_0] {$s_1$}; 
	\node[ state] (s_2) [right=of s_1] {$s_2$}; 
	\node[ state, accepting] (s_3) [right=of s_2] {$s_3$}; 
	
	\path[->] 
	(s_0) edge  node[] {$A$} (s_1)
	edge [loop above, arrow style right] node[align=left]  {$B|C$} ()
	(s_1) edge  node[] {$B$} (s_2)
	edge [loop above] node[ align=left] {$A|C$} ()
	(s_2) edge  node[] {$C$}(s_3)
	edge [loop above] node[ align=left] {$A|B$} ()	
	;
	\node [below right= 0.7cm and 0.8cm of s_1] {State machine of pattern $q = seq~(A;B;C).$};
\end{scope}

\path[big font] (\xEnd -0.4, \yStart) node[above] {$A_{0}$}  ++ (\xEventShift, 0) node[above] {$A_{1}$} ++ (\xEventShift, 0) node[above] {$B_{2}$} ++ (\xEventShift, 0) node[above] {$B_{3}$} ++ (\xEventShift, 0) node[above] {$A_{4}$}++ (\xEventShift, 0) node[above] {$C_{5}$}++ (\xEventShift, 0) node[above] {$B_{6}$} ++ (\xEventShift, 0) node[above= 3pt] {$...$};

\draw [->, color=blue, line width=1pt, big font] (\xEnd + \xEventShift*6.5, \yStart + 1.5) node[above= -3pt, align=center,  color=black, big font] {recent event} -- (\xEnd + \xEventShift*7 + 0.4, \yStart + 0.75);
\draw [<-, line width=1pt] (\xStart, \yStart) -- (\xEnd, \yStart)  node [pos= 0.05, above= 2.5pt , align=left, name=txt] {input event \\stream ($S_{in}$)};
\node[below= 0.8cm of txt.west,anchor=west] {time};

\path[] (\xEnd  - 0.4, \yStart-1.25) node[above] {$A_0$}  ++ (\xEventShift, 0) node[above] {$A_1$} ++ (\xEventShift, 0) node[above] {$B_2$} ++ (\xEventShift, 0) node[above] {$B_3$} ++ (\xEventShift, 0) node[above] {$A_4$}++ (\xEventShift, 0) node[above] {$C_5$}++ (\xEventShift, 0) node[above] {$B_6$} ++ (\xEventShift, 0) node[above= 3pt] {$...$};

\draw [color= w1Color, line width=1pt]  (\xEnd -7,\yStart +0.5 - 1.25) node [below left= 2pt] {$w_1$} -- (\xEnd  -7,\yStart  - 1.25) -- (\xEnd ,\yStart - 1.25)  -- (\xEnd , \yStart + 0.5 - 1.25) ; 

\path[big font] (\xEnd  - 0.4 + \xEventShift*2, \yStart-2.25) node[above] {$B_0$} ++ (\xEventShift, 0) node[above] {$B_1$}
++ (\xEventShift, 0) node[above] {$A_2$}++ (\xEventShift, 0) node[above] {$C_3$}++ (\xEventShift, 0) node[above] {$B_4$} ++ (\xEventShift, 0) node[above= 3pt] {$...$};

\draw [color= w2Color, line width=1pt] (\xEnd  -8,\yStart + 0.5 - 2.25) node [below left= 2pt] {$w_2$} -- (\xEnd - 8,\yStart  - 2.25) -- (\xEnd + \xEventShift*2,\yStart - 2.25) -- ( \xEnd  + \xEventShift*2, \yStart + 0.5 - 2.25); 

\path[big font] (\xEnd  - 0.4 + \xEventShift*4, \yStart-3.25) node[above] {$A_0$}++ (\xEventShift, 0) node[above] {$C_1$}++ (\xEventShift, 0) node[above] {$B_2$} ++ (\xEventShift, 0) node[above= 3pt] {$...$};

\draw [color= w3Color, line width=1pt] (\xEnd -9,\yStart + 0.5 - 3.25) node [below left= 2pt] {$w_3$} -- (\xEnd  -9,\yStart  - 3.25) -- (\xEnd  + \xEventShift*4,\yStart - 3.25)  -- (\xEnd + \xEventShift*4, \yStart + 0.5 - 3.25);

\node [] at (\xStart + 4.5 ,-10.2) {$(a)$};

\begin{scope}[node pm style, xshift=10cm, yshift=0cm]
	\draw[color= w1Color, line width=1pt, rounded corners] (-17.25,0.75) rectangle (-9.5, - 3.75)node[above left= 3pt and 3pt ] {$w_1$};	
	\node[] at (-16.75, 0)  {$cplx_1$}; 
	\node[ state]  (s_110) at (-16, 0)  {$s_0$}; 
	\node[ state] (s_111) [right=of s_110] {$s_1$}; 
	\node[ state] (s_112) [right=of s_111] {$s_2$}; 
	\node[ state, accepting] (s_113) [right=of s_112] {$s_3$}; 
	
	\path[->] 
	(s_110) edge  node[] {$A_0$} (s_111)
	(s_111) edge  node[] {$B_2$} (s_112)
	(s_112) edge  node[] {$C_5$}(s_113)
	;

	\node[] at (-16.75, -1)  {$\gamma_2$}; 
	\node[ state]  (s_120) at (-16, - 1)  {$s_0$}; 
	\node[ state] (s_121) [right=of s_120] {$s_1$}; 
	\node[ state] (s_122) [right=of s_121] {$s_2$}; 
	
	\path[->] 
	(s_120) edge  node[] {$A_1$} (s_121)
	(s_121) edge  node[] {$B_3$} (s_122)
	;
	
	\node[] at (-16.75, -2)  {$\gamma_3$}; 	
	\node[ state]  (s_130) at (-16, - 2)  {$s_0$}; 
	\node[ state] (s_131) [right=of s_130] {$s_1$};
	\node[ state] (s_132) [right=of s_131] {$s_2$};

	\path[->] 
	(s_130) edge  node[] {$A_4$} (s_131)
	(s_131) edge  node[] {$B_6$} (s_132)	
	;
	
	\node[] at (-16.75, -3)  {$\gamma_4$}; 	
	\node[ state]   at (-16, - 3)  {$s_0$}; 	
	;

	\draw[color= w2Color,  line width=1pt, rounded corners] (-17.25, -4) rectangle (-9.5, - 6.75)node[above left= 3pt and 3pt ] {$w_2$};	
	\node[] at (-16.75, -5)  {$\gamma_1$}; 
	\node[ state]  (s_210) at (-16, - 5)  {$s_0$}; 
	\node[ state] (s_211) [right=of s_210] {$s_1$}; 
	\node[ state] (s_212) [right=of s_211] {$s_2$}; 
	
	\path[->] 
	(s_210) edge  node[] {$A_2$} (s_211)
	(s_211) edge  node[] {$B_4$} (s_212)
	;
	
	\node[] at (-16.75, -6)  {$\gamma_2$}; 
	\node[ state]  at (-16, - 6)  {$s_0$}; 	
	;

	\draw[color= w3Color,  line width=1pt, rounded corners] (-17.25, -7) rectangle (-9.5, - 9.75) node[above left= 3pt and 3pt ] {$w_3$};	
	\node[] at (-16.75, -8)  {$\gamma_1$}; 
	\node[ state]  (s_310) at (-16, - 8)  {$s_0$}; 
	\node[ state] (s_311) [right=of s_310] {$s_1$}; 
	\node[ state] (s_312) [right=of s_311] {$s_2$}; 

	\path[->] 
	(s_310) edge  node[] {$A_0$} (s_311)
	(s_311) edge  node[] {$B_2$} (s_312)
	;
	
	\node[] at (-16.75, -9)  {$\gamma_2$}; 
	\node[ state]  at (-16, - 9)  {$s_0$}; 				
		
	\node [] at (-13.5 ,-10.2) {$(b)$};		
\end{scope}

\end{tikzpicture}}
	\caption{Example 1.}
	\label{fig:example1}
\end{figure*}
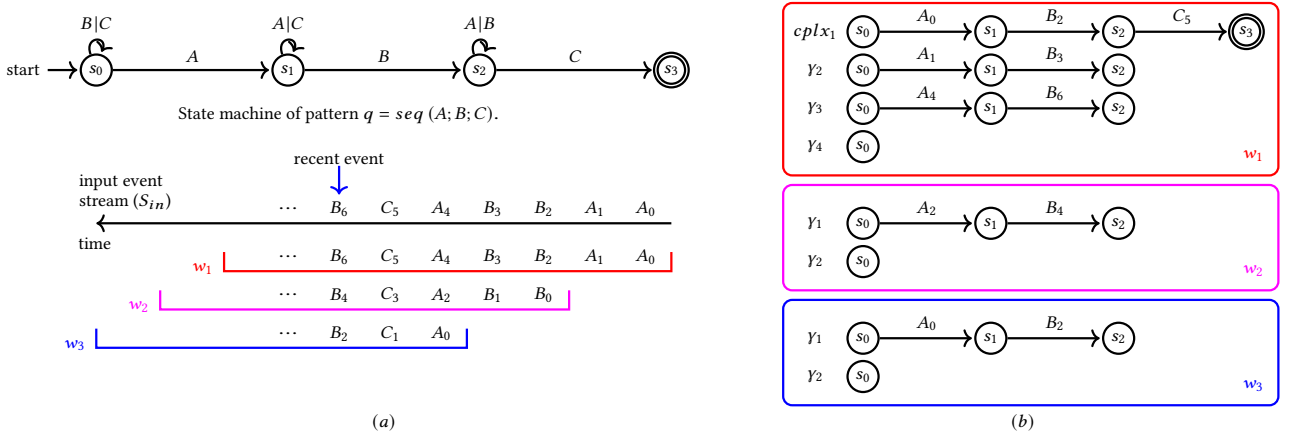

\textbf{\textit{Example 1.}} In a stock application, an operator matches pattern $q$ which correlates stock events from three companies. Pattern $q$ is defined as follows: generate a complex event if a change in the stock quote of company $A$ results in a change in the stock quote of company $B$, followed by a change in the stock quote of company $C$. We may write this pattern as a sequence operator \cite{snoop}: $q= seq (A; B; C)$. Hence, the set of patterns that the operator matches is $Q= \{q\}$. In this example, the event type $T_e$ might represent the company name, i.e., $A$, $B$, and $C$.
Assume that a count-based predicate is used to open windows where a window is opened every two events, i.e., window slide size is two. Figure \ref{fig:example1} depicts this example.  Figure \ref{fig:example1}(a) shows that events in the input event stream ($S_{in}$) are ordered by the sequence number.  Moreover, it shows that there are three open windows which overlap. As an example to show how the same event may have different positions within different windows, we see that the event $A_4$ from the input event stream belongs to all three windows, where it has the positions 4, 2, and 0 within windows $w_1$, $w_2$, and $w_3$, respectively.  

Windows of events are first pushed to the input queue of a CEP operator. The operator continuously gets events from the input queue where, within every window to which an event belongs, the operator checks if the event matches the given pattern(s). We refer to this checking as processing the event within the window. As mentioned above, windows might overlap. However, events within each window are processed independently. 
A pattern in CEP is modeled as a finite state machine \cite{spectre:2017, Ray:2016:SPS:2882903.2882947} (cf. Figure \ref{fig:example1}(a)). The set of all possible states  $\mathbb{S}_{q_i}$ of pattern $q_i \in \mathbb{Q}$ is defined as: $\mathbb{S}_{q_i}= \{s_k : j \leq k < j + m_i\}$, where $m_i$ represents the number of all possible states of pattern $q_i$ and $j$ represents the sum of the  number of all possible states of all patterns $q_l \in \mathbb{Q}$ where $l < i$, i.e.,  $j = \sum_{l = 1}^{i -1} m_l $. In Example 1, pattern $q$ has four states (i.e., $m_i = 4$) where $\mathbb{S}_{q}= \{s_0, s_1, s_2, s_3\}$ as shown in Figure \ref{fig:example1}(a). In the figure, $s_0$ represents the initial state of pattern $q$ and $s_3$ represents its final state.
We define the set of all possible states for all patterns as follows: $\mathbb{S_Q}= \bigcup_{i=1}^{n} \mathbb{S}_{q_i}$. In Example 1, since there is only one pattern (i.e., $\mathbb{Q}= \{q\}$), $\mathbb{S_Q}= \mathbb{S}_{q}= \{s_0, s_1, s_2, s_3\}$. 

Whenever an operator starts to process events within a window, it starts an instance of the state machine of every pattern $q_i \in \mathbb{Q}$ at the initial state. During event processing within a window, an event is matched with the state machine instances of pattern $q_i \in \mathbb{Q}$. The event might cause the state machine instance(s) of pattern  $q_i$ to transit between different states of $\mathbb{S}_{q_i}$. Please recall that we have already defined a partial match. However, let us define it more formally.  An instance of the state machine of pattern  $q_i$ is called a partial match (short PM), where the partial match completes and becomes a complex event if the state machine instance transits to the final state. Hence, processing an event within a window implies that the event is matched with PMs within the window. 
We define a partial match $\gamma$ of pattern $q_i$ as $\gamma \subset q_i$. Moreover, we refer to matching event $e$ with PM $\gamma \in q_i$ as processing event $e$ with PM $\gamma$, denoted by $e \otimes \gamma$. 
In Example 1, assume that the operator matches the events in windows chronologically~\cite{snoop} and the operator has already processed all available events in all open windows, i.e., the operator has processed the last event of type $B$ ($B_6$ in the input event stream) in all windows. Figure \ref{fig:example1}(b)  shows the result of pattern matching in all windows. In window $w_1$, the operator has detected one complex event ($cplx_1$) while there are still three open PMs in window $w_1$: $\gamma_2$, $\gamma_3$,  and $\gamma_4$ . Similarly, there are two PMs in windows $w_2$ and $w_3$ each: $\gamma_1$  and $\gamma_2$.

Partial match $\gamma \subset q_i$ might be at any state of pattern $q_i$ except the final state, where PM $\gamma$ at the final state has already completed and become a complex event. Therefore, the set of all possible states ($\mathbb{S}_{\gamma}$) of  PM $\gamma$ is defined as follows: $\mathbb{S}_{\gamma} = \mathbb{S}_{q_i} \setminus \{\mathit{final~states}\}$.  Hence, the set of all possible states $\mathbb{S}_{\Gamma}$ of all PMs of all patterns is defined as follows: $\mathbb{S}_{\bbGammaVar}= \bigcup_{i=1}^{n} \mathbb{S}_{\gamma_i}: \gamma_i \subset q_i$. In example 1, for PM $\gamma \subset q$, $\mathbb{S}_\gamma= \{s_0, s_1, s_2\}$ and $\mathbb{S}_{\bbGammaVar}= \mathbb{S}_\gamma= \{s_0, s_1, s_2\}$, as there is only one pattern in this example.
We refer to the current state of PM  $\gamma$ as $S_{\gamma}$. Additionally, we refer to PM $\gamma$ at state $s$ as $\gamma_s$.
If processing event $e$  with PM $\gamma \subset q_i$ at state $s$ (i.e., $e \otimes \gamma_s$) causes $\gamma$ to progress, i.e., $e$ matches $q_i$ and causes the state machine instance to transit,  we refer to this as event $e$ contributes to PM $\gamma$ at state $s$, denoted by $e \in \gamma_s$. 
In Example 1, event $B_0$ in window $w_2$ has been processed with $\gamma_1$ at state $s_0$ (i.e., $B_0 \otimes \gamma_{1s_0}$) but it did not cause  $\gamma_1$ to progress. While in the same window $w_2$, event $A_2$ has been processed with $\gamma_1$ at state $s_0$  (i.e., $A_2 \otimes \gamma_{1_{s_0}}$) and it caused $\gamma_1$ to progress to state $s_1$. Hence, event $A_2$ contributes to PM $\gamma_1$ at state $s_0$, i.e., $A_2 \in \gamma_{1_{s_0}}$.
In window $w$, at a certain window position $P$, there might exist one  or more PMs belonging to the same or different patterns $q_i \in \mathbb{Q}$.  We denote the set of PMs that are currently active at window position $P$ by $\bbGammaVar^P_w$. Also, we denote the \textit{total} number of PMs that are opened until the end of window $w$ by $\bbGammaVar^T_w$. In Example 1 Figure\ref{fig:example1}(b), the set of current PMs in windows $w_1$, $w_2$ and $w_3$ are as follows: $\bbGammaVar^6_{w_1} = \{\gamma_2, \gamma_3, \gamma_4\}$, $\bbGammaVar^4_{w_2}= \{\gamma_1, \gamma_2\}$, and $\bbGammaVar^2_{w_3}= \{\gamma_1, \gamma_2\}$. 
Please note that in the negation operator \cite{Ray:2016:SPS:2882903.2882947, Wu:2006:HCE:SASE} if the negated event $e'$ contributes to PM $\gamma$ (i.e., $e' \in \gamma$), PM $\gamma$ is abandoned. For ease of presentation, hereafter, we also refer to the abandoned PMs as completed PMs.

\subsection{Problem Statement}
A CEP operator might have limited resources where, in overload cases, it must perform load shedding by dropping a portion of the input events to avoid violating a given latency bound (LB). However, dropping events might degrade QoR, i.e., resulting in false positives and false negatives. Therefore, the load shedding must be performed in a way that has minimum adverse impact on QoR. 

As we mentioned above, an operator might detect multiple patterns $\mathbb{Q}$ and each pattern has its weight (i.e., $\mathbb{W_Q}$). For pattern $q_i \in \mathbb{Q}$, we define the number of false positives as $FP_{q_i}$ and the number of false negatives as $FN_{q_i}$.
The total number of false positives (denoted by $FP_\mathbb{Q}$) for all patterns is defined as the sum of the number of false positives for each pattern multiplied by the pattern's weight (cf. Equation \ref{eq:fp_Q}). Similarly, the total number of false negatives (denoted by $FN_\mathbb{Q}$) for all patterns is defined as the sum of the number of false negatives for each pattern multiplied by the pattern's weight (cf. Equation \ref{eq:fn_Q}). 
\begin{equation}
\begin{aligned}
\ & \ FP_\mathbb{Q} = \sum_{q_i \in \mathbb{Q}} w_{q_i} . FP_{q_i}
\end{aligned}
\label{eq:fp_Q}
\end{equation}
\begin{equation}
\begin{aligned}
\ & \ FN_\mathbb{Q} = \sum_{q_i \in \mathbb{Q}} w_{q_i} . FN_{q_i}
\end{aligned}
\label{eq:fn_Q}
\end{equation}
As a result, the impact of load shedding on QoR is measured by the sum of the total number of false positives ($FP_\mathbb{Q}$) and the total number of false negatives ($FN_\mathbb{Q}$).
The  objective is to minimize the adverse impact on QoR, i.e., minimize ($FP_\mathbb{Q} + FN_\mathbb{Q}$), while dropping events such that the given latency bound $LB$ is met.  More formally, the objective is defined as follows.
\begin{equation}
\begin{aligned}
minimize \quad  & \ ( FP_\mathbb{Q} +  FN_\mathbb{Q})	\\
\textrm{s.t.} \quad & \ l_e \le LB \quad \forall~ e \in S_{in}
\end{aligned}
\end{equation}
where $l_e$ is the latency of event $e$  that represents the sum of the queuing latency of event $e$ and the time needed to process event $e$ within all windows to which event $e$ belongs.

\vspace{0.3cm}
\section{Load Shedding in CEP}
We extend a CEP operator with our proposed load shedding system (\framework) that in overload cases drops a portion of the input events to maintain the given latency bound (LB). 
In CEP, a load shedding system must perform the following three tasks: 1) deciding when input events must be dropped, 2) computing the time interval and the number of events that must be dropped in every time interval (denoted by drop interval) to maintain LB, and 3) dropping input events that have the lowest adverse impact on QoR. Tasks 1 and 2 have already been well studied in literature \cite{espice, pspice}. Therefore, our focus in this paper is on task 3, i.e, deciding which events to drop. In the following, we shortly explain how tasks 1 and 2 might be performed. Figure \ref{fig:ls} depicts a CEP operator extended with two components to enable load shedding: 1) overload detector and 2) load shedder (LS).

The given latency bound (LB), the rate of incoming input events, and the operator throughput (maximum service rate) can be used as parameters to decide when to drop events. The overload detector periodically monitors these parameters. If the input event rate ($R$) is higher than the operator throughput ($\mu$) for a long enough period, the given latency bound (LB) might be violated. To prevent violating LB, the overload detector requests the load shedder to drop a certain amount  of input events. As a drop interval ($\lambda$), we might use the window size $ws$  or a part of it as proposed in \cite{espice}. Our approach works with any drop interval. However, in this paper, to simplify the presentation, we consider that the drop interval equals the window size, i.e., $\lambda= ws$. 
The number of events that must be dropped in every window to maintain $LB$ can be computed depending on the input event rate $R$ and the operator throughput $\mu$, where the overload detector computes the drop amount $\rho$ per window (i.e., per drop interval) as follows: $\rho= (1- \dfrac{\mu}{R}) . ws$.  After that, the overload detector sends a command containing the drop interval $\lambda$ and the number of events $\rho$ to drop  per $\lambda$ to the load shedder. The load shedder drops $\rho$ events per drop interval $\lambda$ to maintain $LB$.

\begin{figure}[t]
	\centering
	\resizebox{0.7\linewidth}{!}{

\newcommand{\gear}[6]{%
  (0:#2)
  \foreach \i [evaluate=\i as \n using {\i-1)*360/#1 }] in {1,...,#1}{%
   arc (\n:\n+#4:#2) {[rounded corners=1.5pt] -- (\n+#4+#5:#3)
    arc (\n+#4+#5:\n+360/#1-#5:#3)} --  (\n+360/#1:#2)
  }%
  (0,0) circle[radius=#6] 
}

\tikzset{label style/.style={font=\fontsize{24}{26}\selectfont, color=black},
	rectangle style/.style ={color= blue, line width=4pt }, 
	arrow style/.style={color=black, line width=2pt, font=\fontsize{24}{26}\selectfont\bf, -{>[scale=2.5, length=7, width=4]}}, 
	instance style/.style={color=blue, line width=3pt} }

\begin{tikzpicture}[]

\draw  [arrow style, label style ] (2.5, 3) -- (5.5,3) node[pos=0.4, below, align= left] {windows};

\newcommand\XStart{5.5}
\newcommand\YStart{2.5}
\newcommand\SquareSize{1}
\foreach \n in {0,...,4}
	\draw[rectangle style, rounded corners] (\n + \XStart, \YStart) rectangle (\n + \XStart + \SquareSize, \YStart+ \SquareSize);
	
\node [label style ] at (8, 4) {input queue};

\draw  [arrow style, label style ] (10.5, 3) -- (12, 3);
	
\draw [instance style](14, 3 ) ellipse [x radius=3cm,y radius=2.5cm]; 
\begin{scope}[xshift=15.25cm, yshift=3cm]
	\fill[even odd rule]  \gear{8}{0.5}{0.75}{10}{2}{0.25};
\end{scope}
\node[label  style, above=0.55cm] at (15.15, 3) {process};
\node[label  style, above] at (14, 5.5) {operator};
\node[label style, rectangle, draw, blue, line width=2pt, minimum width=0.5cm,minimum height=0.5cm, text= black] (PMs) at (15.15, 1.75) {PMs}; 
\draw [rectangle style, red] (12, 3.75) -- (13.5, 3.25) -- (13.5, 2.75) -- (12, 2.25) --cycle;
\path (13.5, 2.75) -- (12, 2.25) coordinate[pos=0.5] (LSMidBelow);
\path (12, 3.75) -- (13.5, 3.25) coordinate[pos=0.5] (LSMidAbove);
\node[label  style] at (12.75, 3) {LS};
\draw  [arrow style, label style ] (13.5, 3) -- (14.5, 3);
\draw  [arrow style, label style ] (16, 3) -- (20, 3) node[pos=0.7, above=4pt, align= left] {complex\\ events};

\draw[rectangle style] (4.9,-1.5) rectangle (8.5, 0.5)  node [pos=.5,  label style, align=left]{overload\\detector};
\draw[arrow style, label style] (6.75, 2.5)-- (6.75, 0.5);
\draw  [arrow style, label style ] (8.5, -0.5) -- (12.75, -0.5) node[pos=0.6, align= left, below=2pt] {commands};
\draw[arrow style, label style] (12.75, -0.5) -- (LSMidBelow) ;



\end{tikzpicture}}
	\caption{The \framework~Architecture.}
	\label{fig:ls}
	\vspace{-0.3cm}
\end{figure}

\bigskip
\begin{center}
\textbf{\LARGE \framework}
\end{center}

During overload, to maintain the given latency bound (LB), \framework~ drops input events that have the lowest adverse impact on QoR, i.e, on the number of false positives and negatives. To do that,  \framework~ assigns utility values  to the events where an event that has a high impact on QoR has a high utility and vice versa. 
On a high abstraction level, \framework~ works as follows.
1) As mentioned above, an event in a window is processed with PMs within the window. Therefore, in a window, \framework~ assigns utility values to an event for each PM within the window individually, i.e., the event gets a certain utility value for each PM within the window.
2) \framework~ performs load shedding by dropping \textit{events} from \textit{partial matches} within windows. In a window, dropping event $e$ from PM $\gamma$ means that \framework~ prevents processing event $e$ with PM $\gamma$ within the window. 

\framework, primarily, performs two tasks: 1) model building and 2) load shedding. In the model building task, \framework~ predicts the event utilities and summarizes the event utilities to reduce the degradation in QoR in overload situations. In the load shedding task, \framework~ drops events to avoid violating the given latency bound. The model building task is not time-critical and can afford to be heavyweight. On the other hand, the load shedding task is time-critical and hence must be lightweight. 
In the next sections, we describe the above tasks in detail. First, we describe how the utility of an event for a partial match is defined. Then, we explain the way \framework~ predicts the event utility using a probabilistic model. After that, we describe how \framework~ computes the number of events to drop per partial match within windows to maintain the given latency bound. To perform load shedding efficiently, we explain how to predict a utility value that can be used as a threshold utility to drop the required number of events from PMs. Finally, we describe the functionality of the load shedder in \framework.

\subsection{Event Utility}
In a window, only some PMs might complete and become complex events. Hence, PMs in a window might have different importances, w.r.t. QoR.  If a PM completes, it is an important  PM for QoR. Otherwise, it has no impact on QoR.
Moreover, as mentioned above, an event might be processed with one or more PMs within a window, where the event might contribute only to some of these PMs. An event that contributes to a PM might be an important event for the PM since dropping the event from the PM might hinder the PM completion and hence adversely impact QoR. On the other hand, an event that does not contribute to a PM is not important for the PM since dropping the event from the PM does not influence its completion. Therefore, for different PMs in a window, an event might have different importances.
As a result, in  a window, for event $e$ and PM $\gamma$ within the window, \framework~  assigns a utility value to event $e$ (denoted by the utility of event $e$ for PM $\gamma$) depending on the importance of PM $\gamma$ in the window and on the importance of event $e$ for $\gamma$. Higher is the importance of $\gamma$ in the window and higher is the importance of event $e$ for $\gamma$, higher is the utility of event $e$ for $\gamma$.

The utility of event $e$ for PM $\gamma$ of pattern $q_i \in \mathbb{Q}$ within a window (denoted by $U_{e,\gamma}$) depends on three factors: 1) contribution probability---the probability that event $e$ contributes to PM $\gamma$, i.e., $e \in \gamma$, 2) completion probability---the probability that PM $\gamma$ completes,  and 3) pattern weight $w_{q_i}$ (given by a domain expert).  Clearly, if event $e$ has a high probability to contribute to PM $\gamma$, event $e$  is an important event for PM $\gamma$. We consider the completion probability of a PM in computing the event utility as well since the PM is only useful if it completes.
Therefore, if event $e$ has a high probability to contribute to PM $\gamma$ and $\gamma$ has a high probability to complete, event $e$ is an important event and should be assigned a high utility value. This is because dropping event $e$ may hinder PM $\gamma$ to complete and hence it may adversely impact QoR. 

As a result, the utility $U_{e,\gamma}$ of event $e$ for PM $\gamma \subset q_i$ within a window depends on the pattern weight $w_{q_i}$ and the following probability:  $P(e \in \gamma ~\cap~ \gamma ~completes)$, i.e., the probability that PM $\gamma$ completes and event $e$ contributes to PM $\gamma$.
In window $w$, to predict $P(e \in \gamma ~\cap~ \gamma ~completes)$  and hence $U_{e,\gamma}$,   \framework~  uses three features: 1) current state $S_{\gamma}$ of PM $\gamma$, 2) event type $T_e$, and 3) position $P_e$ of event $e$ in window $w$. Therefore, the utility $U_{e,\gamma}$ of event $e$ for PM $\gamma$ of pattern $q_i$ (i.e., $\gamma \subset q_i$) is defined as a function (called utility function) of these three features as shown in Equation \ref{eq:u_pm}:
\begin{equation} 
U_{e, \gamma}= f(T_e, P_e, S_{\gamma})= w_{q_i} . P(e \in \gamma ~\cap~ \gamma ~completes)
\label{eq:u_pm}
\end{equation}
The current state  $S_{\gamma}$ of PM $\gamma$ determines which event type(s) enables PM $\gamma$ to progress, i.e., to transit to a new state(s). Therefore, those two features, i.e., current state $S_{\gamma}$ of the PM and event type $T_e$ are important features for computing $U_{e, \gamma}$. 
For instance, in Example 1,  PM $\gamma$ at state $s_0$ (i.e., $\gamma_{s_0}$), might transit to state $s_1$ only if event $e$ of type $T_e= A$ is processed with PM $\gamma$ (i.e., $e \otimes \gamma_{s_0}$).

The position $P_e$ of event $e$ in window $w$ is an important feature to compute $U_{e, \gamma}$ as well since it determines the number of remaining events in the window. If there are still many events remaining in a window, the probability of a PM  to complete might be higher than the case where there are only a few remaining events in the window. This is because, in case of many remaining events in a window, a PM  has a chance to be processed with more events than in case of only a few remaining events in the window and hence the PM has a higher chance to progress.
Moreover, the event position $P_e$ represents the temporal distance between events within the same window. It determines which event instance(s) of the same event type has a higher probability to contribute to a PM in the window as shown in \cite{espice}. This is because there exists a correlation between events of certain types at certain positions within a window. A change in an event of a certain type influences the change of events of other types within a certain time interval, i.e., certain position(s) within the window. In Example 1, in  window $w$, a change in the stock quote of company $A$, i.e., $T_e= A$,  at a certain point of time $t_1$ (i.e., at a certain position in window), might cause a change in the stock quote of company $B$, i.e., $T_e= B$, within a certain time interval $\left]t_1, t_2 \right]$, i.e., within certain position(s) in the window.

\definecolor{LightCyan}{rgb}{0.88,1,1}
\newcolumntype{g}{>{\columncolor{LightCyan}}c}
\renewcommand{\arraystretch}{1}
\setlength{\fboxrule}{1pt}
\begin{figure}	
	\fbox{
\begin{minipage}[t]{1.03\linewidth}	
		\centering	
	\begin{minipage}[t]{\linewidth}	
		\begin{minipage}[t]{.40\linewidth}
			\centering
			\resizebox{0.99\linewidth}{!}{

\definecolor{darkGreen}{rgb}{0,0.5,0}
\tikzset{label style/.style={font=\huge\bf, color=black},
	rectangle style/.style ={color= blue, line width=2pt }, 
	arrow style/.style={color=black, line width=2pt, font=\Large, -{>[scale=2.5, length=7, width=4]}}, 
	instance style/.style={color=blue, line width=3pt},
	line style/.style={ line width=1.25pt, font=\Large},
	node pattern style/.style= {shorten >=1pt, node distance=1.5cm,on grid,auto, line width=1pt, font=\Large},	
	node pm style/.style= {shorten >=1pt, node distance=2 cm,on grid,auto, line width=1pt, font=\Large},
	every state/.style={minimum size=0pt, inner sep=2pt}	
}

\begin{tikzpicture}[] 

\begin{scope}[node pattern style]
	\node[ state,initial]  (s_0) at (0, 0)  {$s_0$}; 
	\node[ state] (s_1) [right=of s_0] {$s_1$}; 
	\node[ state, accepting] (s_2) [right=of s_1] {$s_2$}; 
	
	\path[->] 
	(s_0) edge  node[] {$A$} (s_1)
	edge [loop above] node[align=left]  {$B$} ()
	(s_1) edge  node[] {$B$} (s_2)
	edge [loop above] node[ align=left] {$A$} ()
	;
\end{scope}

\end{tikzpicture}}
			\caption*{State machine for pattern $q= seq~(A;B)$.}
		\end{minipage} %
		\hfill
		\begin{minipage}[t]{.55\linewidth}
			\centering
			\begin{tabular}{| g | c| c | c |c |c |c |}
				\hline 
				\rowcolor{LightCyan}
				$T_e / P_e$ & 0 & 1 & 2 & 3 & 4  \\ \hline
				$A$ & x & x & x & x &  \\ \hline
				$B$ &  &  & x & x & x \\ 
				\hline
			\end{tabular}
			\captionof{table}{Event distribution within windows.}
			\label{tab:event-distribution}
		\end{minipage} %
	\end{minipage}
	\hfill
	\begin{minipage}[t]{\linewidth}
		\begin{minipage}[t]{.35\linewidth}
			\centering
			\begin{tabular}[]{ | g | l |}
				\hline
				\multirow{1}{*}{$A_0$} &$ob_e \langle 1-2, s_0, s_1, A_0 \rangle: \frac{2}{6}$ \\ \hline   
				
				
				\multirow{1}{*}{$A_2$} &$ob_e \langle 3-4, s_0, s_1, A_2 \rangle: \frac{2}{4}$ \\ \hline  
				
				\multirow{1}{*}{$A_3$} &$ob_e \langle 5-6, s_0, s_1, A_3 \rangle: \frac{2}{2}$ \\ \hline  
				
			\end{tabular}
		\end{minipage} 
		\hfill
		\begin{minipage}[]{.48\linewidth}
			\begin{minipage}[t]{\linewidth}
				\begin{tabular}{ | g | l |}
					\hline
					\multirow{1}{*}{$B_3$} &$ob_e \langle 1, s_1, s_2, B_3 \rangle: \frac{1}{4}$ \\ \hline  
					
					\multirow{1}{*}{$B_4$} &$ob_e \langle 2- 3, s_1, s_2, B_4 \rangle: \frac{2}{5}$ \\ \hline  
					
					\hline
				\end{tabular}
			\end{minipage} 
			\hfill \vspace{0.1cm}
			\begin{minipage}[]{\linewidth}
				\begin{tabular}{| l |}
					\hline
					$ob_{\gamma} \langle 1-3, ~completed \rangle$ \\ 	\hline		
					$ob_{\gamma} \langle 4-6, ~not~ completed \rangle$ \\
					
					\hline
				\end{tabular}
			\end{minipage} 
			
		\end{minipage}
		\captionof{table}{Contribution $ob_e$ and completion $ob_\gamma$ observations.}
		\label{tab:obsevations}
	\end{minipage}
\end{minipage}	
}
	\caption{Observations gathered from six PMs.}
	\label{fig:observations-example}
\end{figure}

\renewcommand{\arraystretch}{1}

\definecolor{Gray}{gray}{0.95}
\renewcommand{\arraystretch}{1}
\setlength{\fboxrule}{1pt}
\begin{figure}
	\centering
	\fbox{
\begin{minipage}[t]{1\linewidth}	
	\begin{minipage}[t]{.95\linewidth}	
		\begin{minipage}[t]{.45\linewidth}
			\centering
			\begin{tabular}{| g | c| c | c |c |c |c |}
				\hline 
				\rowcolor{Gray}
				\multicolumn{6}{|c|}{$s_0$} \\ \hline	
				\rowcolor{LightCyan}
				$T_e / P_e$ & 0 & 1 & 2 & 3 & 4 \\ \hline
				$A$ & 33 & 0 & 25 & 0 & 0 \\ \hline
				$B$ & 0 & 0 & 0 & 0 & 0 \\ 
				\hline
			\end{tabular}
		\end{minipage} 
		\hfill
		\begin{minipage}[]{.45\linewidth}
			\centering
			\begin{tabular}{|g| c| c | c |c |c |c |}
				\hline
				\rowcolor{Gray}				
				\multicolumn{6}{|c|}{$s_1$} \\ \hline
				\rowcolor{LightCyan}
				$T_e / P_e$ & 0 & 1 & 2 & 3 & 4 \\  \hline
				$A$ & 0 & 0 & 0 & 0 & 0 \\  \hline
				$B$ & 0 & 0 & 0 & 25 & 40 \\ 
				\hline
			\end{tabular}
		\end{minipage}
	\end{minipage}

\end{minipage}
}
	\caption{Computing event utility  $U_{e,\gamma}$ for a partial match.}
	\label{fig:utility-example}

\end{figure}

\renewcommand{\arraystretch}{1}

\subsection{Predicting Event Utility}
Having defined the utility $U_{e, \gamma}$ of event $e$ for PM $\gamma$, now, we describe how \framework~ predicts the utility $U_{e,\gamma}$ within a window, i.e.,  $P(e \in \gamma ~\cap~ \gamma ~completes)$, hence predicting the value of utility function $f(T_e, P_e, S_{\gamma})$ in Equation \ref{eq:u_pm}.  
For ease of presentation, we introduce a simple running example which is depicted in Figures \ref{fig:observations-example} and \ref{fig:utility-example}.\\
\textbf{\textit{Example 2}}. Let us assume that an operator matches a pattern $q= seq~(A;B)$, where $\mathbb{S}_{q}= \{s_0, s_1, s_2\}$ and $\mathbb{S}_{\gamma}= \{s_0, s_1\}$, $\gamma \subset q$. The used  window length is 5 events (i.e., $ws= 5$) and there are only two event types in the input event stream: $A$ and $B$.
 
To predict the utility $U_{e,\gamma}$ of event $e$ for PM $\gamma$ of pattern $q_i$ in window $w$,  we first need  to predict the \textit{completion probability} of PM $\gamma$, i.e., find the probability that  PM $\gamma$ at state $S_{\gamma}$ and at position $ P_e$ in window $w$ will complete. Additionally, we need to predict the \textit{contribution probability} of event $e$ to PM $\gamma$, i.e., the probability that event $e$ of type  $T_e$ at position $P_e$ in window $w$ contributes to PM $\gamma$ ($e \in \gamma$).
If the contribution and completion probabilities are high, then the event utility $U_{e, \gamma}$ is high. On the other hand, if the contribution  and/or completion probabilities are low,  then the event utility $U_{e, \gamma}$ is low. 
\framework~ uses statistics gathered over already processed windows to predict the completion and contribution probabilities, thus predicting the event utility for PMs. Next, we first show which statistics \framework~ gathers. Then, we explain the way the event utility $U_{e, \gamma}$ for PMs is predicted depending on those gathered statistics.

\textbf{Statistic Gathering. }
To predict the contribution and completion probabilities  (i.e., to predict $P(e \in \gamma ~\cap~ \gamma ~completes)$), thus predicting the value of utility function $f$, \framework~ gathers statistics on the progress of PMs within windows during event processing in an operator. To do that, \framework~ uses two types of \textit{observations}: 1) contribution observation, denoted by $ob_{e}$, and 2) completion observation, denoted by $ob_{\gamma}$. In  window $w$, for each event $e$ within $w$,  whenever event $e$ is processed with PM $\gamma$ at state $s=S_{\gamma}$ (i.e, $e \otimes \gamma_s$), the operator builds an observation of type contribution  $ob_e \langle id, s, s', e \rangle$, where $id$ is the $id$ of PM $\gamma$. $s'$ represents the state of PM $\gamma$ after processing event $e$. If $s \neq s'$, event $e$ has contributed to PM $\gamma$ at state $s$, i.e., $e \in \gamma_s$.  Additionally, in window $w$, if PM $\gamma$ completes, the operator builds an observation of type completion $ob_{\gamma} \langle id, completed \rangle$, where again $id$ is the id of PM $\gamma$. When window $w$ closes ( i.e., all its events are processed), all still open PMs in window $w$, i.e., $\bbGammaVar^P_{w}$, (here $P$ is the last position in $w$) are considered as \textit{not completed} PMs. 

Figure \ref{fig:observations-example} shows an example of gathered observations on six PMs. Table \ref{tab:event-distribution} shows the distribution of event types in different positions within a window where a cell with $x$ sign in the table means that the corresponding event type might be present at the corresponding position within a window.  Please note that   event types might not be present in all positions within a window. In the table, for example, the event type $A$ never comes at position 4 in any window and event type $B$ does not come at positions 0 and 1 in any window. Table \ref{tab:obsevations} shows observations on event $e$ of type $T_e$ at position $P_e$ in a window and PM $\gamma$ at state $s$ only if $e$ contributes to $\gamma$ (i.e., $e \in \gamma_s$).
For example, in the table, event $B_3$ of type $T_e= B$ at position $P_e=3$ within windows has never contributed to PM $\gamma$ at state $s_0$. Therefore, there are no observations shown in the table on event $B_3$ with a PM at state $s_0$.
Clearly, if event $e$ is not present at a certain position within windows, event $e$ can not contribute to any PM at this window position. For example, as shown in Table \ref{tab:event-distribution}, the event of type $B$ never comes at position $1$ within windows. Therefore, there are  no observations on the event type $B$ at position $1$ within windows with a PM at any state. In Table \ref{tab:obsevations}, next to each observation of type contribution $ob_e$, we show the number of PMs at state $s$ to which an event \textit{contributed} divided by the \textit{total}  number of PMs at state $s$ with which an event is processed, i.e., $\frac{|\{e: e \in \gamma_s\}|}{|\{e: e \otimes \gamma_s\}|}$. For example, in the table, $ob_e \langle3-4, s_0, s_1, A_2\rangle: \frac{2}{4}$ means that the event of type $T_e= A$ at position 2 within windows has been processed with four PMs at state $s_0$. However, it has contributed only to two PMs, in particular, it has contributed to  PMs 3 and 4. The table also shows which PMs have completed. For example, in the table, PMs $\gamma_1$, $\gamma_2$, and $\gamma_3$ have completed while PMs $\gamma_4$, $\gamma_5$, and $\gamma_6$ have not completed.

After gathering statistics from $\eta$ observations, \framework~ uses these observations  to predict the utility  $U_{e, \gamma}$ of event $e$ for PM $\gamma$ within window $w$, i.e., to predict the utility function $f$ (cf. Equation \ref{eq:u_pm}).

\textbf{Utility Prediction. }
\framework~ uses the gathered observations of both types (contribution $ob_e$ and completion $ob_{\gamma}$) to predict the probability value $P(e \in \gamma ~\cap~ \gamma ~completes)$, hence predicting $U_{e, \gamma}$. First, from both these observation types, \framework~ computes the utility of event $e$ for the set of all possible states of PM $\gamma$ (i.e., $\mathbb{S}_{\gamma}$)  as follows:
\begin{equation}
	   U_{e, s}= \dfrac{| \{e: e \in \gamma_s ~\&~ \gamma ~ completed \}|}{|\{e: e \otimes \gamma_s \}|}
\end{equation}
where $U_{e, s}= P(e \in \gamma_s ~\cap~ \gamma ~completes)$. For event $e$ of certain type $T_e$ at certain position $P_e$ within window $w$  and for  PM  $\gamma$ at certain state $s$, $U_{e, s}$ is computed as a ratio between the  number of times PM $\gamma$ \textit{ completes} and event $e$ \textit{contributes} to PM $\gamma$ at state $s$ (i.e., $e \in \gamma_s$) and the \textit{total} number of times event $e$ is processed with PM $\gamma$  at state $s$ (i.e., $e \otimes \gamma_s $).

Figure \ref{fig:utility-example} shows the computed utility values $U_{e, s}$ from  the observations shown in Table \ref{tab:obsevations}. The values are shown as percentage values. The table shows the utility value of event $e$ of type $T_e$ at  position $P_e$ within a window for PMs at states $s_0$ and $s_1$. For example,  in the table, event $e= A_2$ of type  $T_e=A$ at position $P_e=2$ within a window is processed with four PMs at state $s_0$ (PMs 3, 4, 5, and 6). However, it has contributed only to  two PMs ( 3 and 4). Moreover, since only PM 3 completed, we account for the contribution of event $e= A_2$  only to PM 3.  Therefore, in the table, the utility of event type $T_e= A$ at position $P_e=2$ within a window for a PM at state $s_0$ equals to $25\%$, i.e., $U_{e, s_0}=  \frac{1}{4}= 25\%$. The event type $T_e= A$ has never contributed to a PM  at state $s_1$  since only the event type $T_e= B$ may contribute to a PM at state $s_1$. Therefore, the utility of an event of type $T_e=A$ at any position within a window for a PM at state $s_1$ is always zero as shown in the table.  Similarly, the event type $T_e=B$  never contributes to a PM at state $s_0$. Hence, the utility of an event of type $T_e=B$  at any position within a window for a PM at state $s_0$ is always zero.    

The utility values for all states of PM $\gamma$ of pattern $q_i \in \mathbb{Q}$ together  multiplied by   the pattern weight $w_{q_i}$ represent the predicted utility $U_{e, \gamma}$ of event $e$ for PM $\gamma \subset q_i$, where $U_{e,\gamma_s}= f(T_e, P_e, s)= w_{q_i} . U_{e, s}$. 
Now, we need to store these predicted utility values $U_{e, \gamma}$ for all patterns (i.e., for $\mathbb{Q}$) so that, during load shedding, \framework~ can retrieve them. 
To reduce the storage overhead, in case of large window size, we use bins to group event  utilities. Within  window $w$, the utility values of event $e$ of type $T_e$ at several consecutive window positions (i.e., bin size $bs$)  for PM $\gamma_s$ at state $s$  are grouped together by taking the average utility value of this event type $T_e$ over all these positions for PM $\gamma_s$. For ease of presentation, we will use the bin of size $bs= 1$ if not otherwise stated.
To efficiently retrieve the utility values during load shedding, we store  the utilities in a table (called utility table $UT$) of three dimensions ($M ~x~ N ~x~ K$),  where $M$ represents the number of different event types, $N= \frac{ws}{bs}$, and $K$ is the number of all possible states of all PMs of all patterns, i.e., $K= |\mathbb{S}_{\bbGammaVar}|$.   Therefore, the storage overhead of the utility table $UT$ is $O(M . N. |\mathbb{S}_{\bbGammaVar}|)$. Each cell $UT(T_e, P_e, S_{\gamma})$ in the utility table stores the utility value $U_{e, \gamma}$ of event $e$ of type $T_e$ at position $P_e$ within a window for PM $\gamma$ at state $S_\gamma$, i.e.,  $U_{e,\gamma}= f(T_e, P_e, S_{\gamma})= UT(T_e, P_e, S_\gamma)$. Hence, to get the utility $U_{e, \gamma}$ of event $e$ for PM $\gamma$, \framework~ needs to perform only a single lookup in the utility table $UT$. This means that the time complexity to get $U_{e, \gamma}$ is $O(1)$ which considerably reduces the overhead of load shedding.

The input event stream might change over time, hence the predicted utilities of events for PMs might become inaccurate. One way to capture the changes in the input event stream and keep the event utility accurate is by periodically gathering statistics and recomputing the utility value $U_{e,\gamma}$.

\subsection{Drop Amount}
As we mentioned above, to maintain the given latency bound ($LB$) in an overload situation, we must drop $\rho $ events from every window. However, \framework~ drops events from PMs, not from windows, where an event might be dropped from a PM while it is processed with another PM within the same window. Therefore, we must find a mapping between the number of events to drop per window ($\rho$) and the number of events to drop per PM within the window. To do that, let us first define the virtual window.

\textbf{Virtual Window.} The virtual window ($vw$) of window $w$ is a set which contains  triplets  $(e, s, O)$ consisting of event $e$ of type $T_e$ at position $P_e$ within $w$, state $s \in \mathbb{S}_{\bbGammaVar}$, and  the number of occurrences $O > 0$ which represents the number of times event $e$ has been processed with a PM at state $s$ within window $w$. More formally: 
$vw= \{ (e, s, O): \forall ~e \in w,  ~ \forall ~\gamma \in \bbGammaVar^T_{w},~  O=|\{\gamma: e \otimes \gamma_s\}|>0 \}$.
The virtual window $vw$ of  window $w$ contains information on the number of times event $e$ within window $w$ is processed with each distinct state $s$ of a PM in window $w$.  
The virtual window depends on the states of PMs in a window. Therefore, it is only possible to know the  exact virtual window of  window $w$ when all events in window $w$ are processed, i.e., when the set of all PMs $\bbGammaVar^T_w$ and their states in window $w$ are known. However, we need to know the virtual window of window $w$ before processing all events in window $w$ since we use the virtual window to decide how many and which events must be dropped from PMs within window $w$. 

Therefore, \framework~ predicts virtual window $vw$ of window $w$ by gathering statistics from the operator on  already processed windows, denoted by $W_{stat}$. As mentioned above, in different windows, event distribution might be different (cf. Table \ref{tab:event-distribution}). Additionally, the occurrences of PM states at certain window positions might also be different in different windows. Hence, different windows might have different corresponding virtual windows. Therefore, to predict virtual window $vw$ of window $w$, \framework~ first computes virtual window $vw_j$ for each window $w_j$ in the gathered statistics $W_{stat}$, where $j=1, ..,|W_{stat}|$. Then,  \framework~ combines all triplets $(e, s, O)$ from these virtual windows $vw_j$ to construct the virtual window $vw$ by taking the average value for the number of occurrence $O$ of each triplet, i.e., $ vw=\{(e,s,O): e= e_j, s= s_j, O= O + \frac{O_j}{|W_{stat}|}, ~\forall~(e_j, s_j, O_j) \in vw_j\}$.
The size of virtual window $vw$ (denoted by $ws_v$) is computed as the total number of occurrences of each triplet in $vw$ as follows: $ws_v= \sum_{(e, s, O) \in vw} O$. The \textit{virtual window size} represents the \textit{number of times} events  are processed with PMs in a window. Therefore, the average number of times ($avg_O$) an event  is processed with a PM in window $w$ is computed as follows: $avg_O = \frac{ws_v}{ws}$.
For example, if every event is processed with two PMs within window $w$, then the virtual window size $ws_v$ is twice the window size $ws$ (i.e., $ws_v=  2. ws$) and  $ avg_O= 2$.

Dropping an event from window $w$ implies that the event is dropped from the set of all current PMs $\bbGammaVar^P_w$ within  window $w$. Therefore, if $\rho$ events must be dropped from window $w$, it implies that, in total, $\rho_v \approx \rho . avg_O \approx \rho  .\frac{ws_v}{ws}$ events must be dropped from all PMs $\bbGammaVar^T_w$ in window $w$ (from  virtual window $vw$ of window $w$, as a shorthand). Hence, dropping $\rho$ events from a window is similar to dropping $\rho_v$ events from its virtual window. 
One approach to drop $\rho_v $ events from a virtual window (i.e., $\rho_v $ events in total from all PMs in a window) is to drop events equally (for example, equal percentage) from every PM in the window. However, not all PMs in a window have the same importance/same completion probability. Therefore, the drop amount per PM should take into consideration the importance of PMs in the window which in turn minimizes the adverse impact of dropping on QoR. 
Please note that it is not possible to get the utility of all events for all PMs in a window and then sort them. After that, drop those $\rho_v $ events from PMs that have the lowest utilities. The reason for this is that the event utilities for PMs in a window are only known after processing all events in the window. This is because the event utilities depend on the current state of PMs ($\bbGammaVar^P_w$) in the window which is only known after processing the events in the window.
Next, we explain how to drop the required number of events ($\rho_v$) from the virtual window of each window while considering the importance of PMs in the window.

\textbf{Utility Threshold.} The approach is to find a utility value (called utility threshold $u_{th}$) that is used as a threshold value to drop the needed amount of events from virtual window $vw$ of window $w$. For each triplet $(e, s, O)$ in virtual window $vw$, we get the utility value $u = U_{e,\gamma_s} = f(T_e, P_e, s)$ from the utility table $UT$. As the number of occurrences $O$ in the triplet represents the number of times state $s$ might occur at window position $P_e$, the  number of occurrences $O$ implies that the utility value $u= U_{e,\gamma_s}$ might occur $O$ times in virtual window $vw$, denoted by the utility occurrences $O_u$  for utility $u$, i.e, $O_u= O$. We accumulate the number of utility occurrences $O_u$  for all utility values in $vw$ in ascending order, denoted by the accumulative utility occurrences $OC_u$ for the utility $u$,  as follows: $OC_u = \sum_{u' \leq u} O_u'$. The accumulative utility occurrences $OC_u$ for utility $u$ means that there exist $OC_u$ events in virtual window $vw$ which have a utility value less or equal to the utility value $u$.

Therefore, using $u$ as a threshold utility $u_{th}$ enables \framework~ to drop $OC_u$ events from PMs in a window. Hence, to drop $\rho_v$  events from the virtual window, we must find a utility value $u= u_{th}$, where $OC_u= \rho_v$. 
To efficiently retrieve the utility threshold, we store the accumulative utility occurrences in an array (denoted by utility threshold array ($UT_{th}$)) of the same size as the virtual window size $ws_v$ as follows: $UT_{th}(i)= u$, where $i= 1,.., ws_v$ and  $OC_u\ge i$ and $OC_u < OC_{u'} ~ \forall ~u < u'$. Therefore, to drop $\rho_v$  events from the virtual window, $u_{th}= UT_{th}(\rho_v)$. Hence, the time complexity to get $u_{th}$ is $O(1)$. Please note that predicting the virtual window and building the utility threshold array are done during the model building task. While during the load shedding, \framework~ performs the following two tasks that have a time complexity of $O (1)$: 1) computing how many events to drop (i.e., $\rho_v$) per virtual window, and 2) determining what utility threshold (i.e., $u_{th}$) to use.

\subsection{Load Shedding}
In the above sections, we showed how to compute the utility of events for PMs within a window and how to predict the utility threshold. Now, we describe how \framework~ performs the load shedding, i.e., deciding whether an event should be dropped from a PM or not. Algorithm \ref{alg:loadShedding} clarifies how load shedding is performed.

For each event $e$ within window $w$, before processing $e$ with  PM $\gamma$ in window $w$, the operator asks the load shedder (LS) whether to drop event $e$ from PM $\gamma$. If the LS returns True, the operator drops event $e$ from PM $\gamma$, otherwise, it processes event $e$ with PM $\gamma$.
If there is no overload on the operator, there is no need to drop events and hence LS returns False which means that the operator can process event $e$ with PM $\gamma$ (cf. Algorithm \ref{alg:loadShedding}, lines 2-3). On the other hand, if there is an overload on the operator, LS checks whether the utility $U_{e, \gamma}$ of event $e$ for PM $\gamma$ is higher than the utility threshold $u_{th}$. Therefore, the LS first gets the utility $U_{e, \gamma}$ of event $e$ for PM $\gamma$ from the utility table $UT$, where $U_{e,\gamma}= f(T_e, P_e, S_\gamma)= UT(T_e, P_e, S_\gamma)$.  After that, \framework~ compares the utility value with the utility threshold $u_{th}$, where it returns True if $U_{e,\gamma} \leq u_{th}$, otherwise \framework~ returns False (cf. Algorithm \ref{alg:loadShedding}, lines 4-7). This shows that \framework~ is lightweight in performing load shedding where the time complexity to decide whether  or not to drop an event from a PM is $O(1)$.

\begin{algorithm}
	\setbox0\vbox{\small
		{\fontsize{8.0}{9.0}\selectfont
			\begin{algorithmic}[1]
				\algsetblockdefx[function]{func}{endfunc}{}{0.2cm}[3]{#1 \textbf{#2} (#3) \textbf{begin}}{\textbf{end function}}				

				\func {}{drop}{$T_e, P_e, S_{\gamma}$}
					\If{$!\mathit{isOverloaded}$} {\fontsize{7.0}{8.0}\selectfont \Comment there is no overload hence no need to drop events}	
						\State $\mathbf{return} \quad False$
					
					\ElsIf {$UT(T_e, P_e, S_{\gamma}) \le \mathit{u_{th}}$}
							\State $\mathbf{return} \quad True$
						\Else
							\State $\mathbf{return} \quad False$	
					\EndIf		
				
				\endfunc
				
			\end{algorithmic}
		}
	}
	\centerline{\fbox{\box0}}
	\caption{Load shedder.}
	\label{alg:loadShedding}
\end{algorithm}
\vspace{-0.3cm}

\section{Performance Evaluations}
\label{sec:results}

In this section, we evaluate the performance of \framework~ by using two real-world datasets and several representative queries.
\subsection{Experimental Setup}

\textbf{\textit{Evaluation Platform.}}
We run our evaluations on a machine that is equipped with 8 CPU cores (Intel 1.6 GHz) and a main memory of 24 GB. The  OS used is CentOS 6.4. We run a CEP operator in a single thread on this machine, where this single thread is used  as a resource limitation. Please note, the resource limitation can be any number of threads/cores and the behavior of  \framework~ does not depend on a specific limitation.
We implemented \framework~ by extending a prototype CEP framework that is implemented using Java.

\textbf{\textit{Baseline.}}
We compare the performance of \framework~ with three state-of-the-art load shedding strategies: 1) eSPICE: it is a black-box load shedding approach that drops events from windows \cite{espice}. 2) BL: we also implemented a black-box load shedding strategy (denoted by BL) similar to the one proposed in \cite{He2014OnLS}.  Additionally,  it captures the notion of weighted sampling techniques in stream processing \cite{3:Tatbul:2003:LSD:1315451.1315479}. BL drops events from windows, where an event type (e.g., player ID or stock symbol) receives a higher utility proportional to its repetition in patterns and in windows. 
Then, depending on event type utilities, it uses uniform sampling to decide which event instances to drop from the same event type. 3) pSPICE: it is a white-box load shedding strategy that drops PMs \cite{pspice}.

\textbf{\textit{Datasets.}}
We use two real-world datasets. 
1) A stock quote stream from the New York Stock Exchange, which contains real intra-day quotes of different stocks from NYSE collected over two months from Google Finance \cite{google_finance}. 
2) A position data stream from a real-time locating system (denoted by RTLS) in a soccer game \cite{debs2013}. Players, balls, and referees  are equipped with sensors that generate events containing their position, velocity, etc. 

\textbf{\textit{Queries.}}
We apply four queries ($Q_1$, $Q_2$, $Q_3$, and $Q_4$) that cover an important set of operators in CEP as shown in Table \ref{tab:queries}: sequence operator, sequence operator with repetition, sequence with negation operator, and sequence with any operator, all with skip-till-next/any-match \cite{Balkesen:2013:RRI:2488222.2488257,  Wu:2006:HCE:SASE, snoop}. Moreover, the queries use time-based sliding window strategy.

In Table \ref{tab:queries}, we use $ws$ to refer to the window length. For stock queries ($Q_1$, $Q_2$, $Q_3$), $C_i$ represents the stock quote of company $i$. $Q_1$ detects a complex event  when rising or falling stock quotes of 10 certain stock symbols, by a given percentage,  are detected within $\mathit{ws}$ minutes in a certain sequence.
Q2 detects a complex event when 10 rising or 10 falling stock quotes of certain stock symbols \textit{with repetition}, by a given percentage,  are detected within $\mathit{ws}$ minutes in a certain sequence.
$Q_3$ is similar to $Q_1$ but it detects a complex event only if the stock quote of a certain company (i.e., $C_5$) does not change by a given percentage. 
$Q_4$ uses the RTLS dataset and it detects a complex event when any 3 defenders of a team (defined as $\mathtt{D_i}$) defend against a striker (defined as $\mathtt{S}$) from the other team within $ws$ seconds from the ball possessing event by the striker. The defending action is defined by a certain distance between the striker and the defenders. For this query, we use two strikers, one from each team.

\begin{table}
\centering
\resizebox{\linewidth}{!}{%
\begin{tabular}{| l | l |}
	\noalign{\hrule height 1pt}
	\multicolumn{2}{!{\vrule width 1pt}c!{\vrule width 1pt}}{\textbf{Stock queries}} \\
	\noalign{\hrule height 1pt}
	$Q_1$ &  \makecell[cl]{\textbf{pattern} $\mathbf{seq} (C_1; C_2;..;C_{10})$ \\
							\qquad $\mathbf{where}~ all~ C_i ~rise~ by~ x\%$ 
							 $\textbf{or} ~ all~ C_i ~fall ~ by~ x\%,~ i= 1.. 10$\\
							\qquad\textbf{within} \textit{ws} \text{minutes} 
							}
							\\ \hline
	
	$Q_2$ &  \makecell[cl]{\textbf{pattern} $\mathbf{seq} (C_1;  C_1; C_2; C_3; C_2; C_4; C_2;$
		$C_5; C_6; C_7; C_2; C_8; C_9; C_{10})$\\
			\qquad $\mathbf{where}~ all~ C_i ~rise~ by~ x\%$ 
			$\textbf{or} ~ all~ C_i ~fall~ by~ x\%,~ i= 1.. 10$\\
			\qquad\textbf{within} \textit{ws} \text{minutes} 	 
		}\\ \hline
	
	$Q_3$ &  \makecell[cl]{\textbf{pattern}  $\mathbf{seq}(C_1;  C_2; C_3; C_4; \mathbf{!C_5}; C_6; C_7; C_8; C_9; C_{10})$   \\                
											\qquad $\mathbf{where}~ all~ C_i ~rise~ by~ x\% ~\mathbf{and}~ C_5 ~does~not~ rise~ by~ y\%$ \\
											\qquad$\textbf{or} ~ all~ C_i ~fall~ by~ x\% ~\mathbf{and}~ C_5 ~does~not~ fall~ by~ y\%$\\
											\qquad \quad $,~ i= 1.. 10 ~ and~ i\ne 5$\\
											\qquad\textbf{within} \textit{ws} \text{minutes} 
		}\\ 
		\noalign{\hrule height 1pt}
	\multicolumn{2}{!{\vrule width 1pt}c!{\vrule width 1pt}}{\textbf{Soccer queries}} \\ 
    	\noalign{\hrule height 1pt}
	$Q_4$ &   \makecell[cl]{\textbf{pattern} $\mathbf{seq} (S; \mathbf{any}(3, D_1, D_2, ..,D_n))$ \\
					\qquad $\mathbf{where}~ S ~possesses ~ball~ \mathbf{and} ~ distance(S, D_i) \leq x~ meters $ \\
						\qquad \qquad $,~ i= 1.. n$  and $n$ is the number of players in a team
						\\
					\qquad\textbf{within} $ws$ \text{seconds}
		}\\  \hline

\end{tabular}
}
\captionof{table}{Queries.}
\label{tab:queries}
\end{table}

\subsection{Experimental Results}
\label{sec:exp_results}
In this section, we evaluate the performance of \framework~ in comparison with other load shedding strategies. First, we show its impact on QoR, i.e., the number of false negatives and the number of false positives. Then, we show how good \framework~is  in maintaining the given latency bound ($LB$).

If not stated otherwise, we use the following settings. 
For all queries $Q_1$, $Q_2$, $Q_3$, and $Q_4$,  we use a \textit{time-based} sliding window and a \textit{time-based} predicate. 
A new window is opened for $Q_1$, $Q_2$, $Q_3$ every 1 minute, i.e., the slide size is 1 minute. For $Q_4$, a new window is opened every 1 second.
We stream events to the operator from the datasets that are stored in files. We first stream events at input event rates which are less or  equal to the operator throughput $\mu$ (maximum service rate) until the model is built. After that, we increase the input event rate to enforce load shedding as we will mention in the following experiments. 
The used latency bound $LB = 1$ second. We configure all load shedding strategies (i.e., \framework, eSPICE, BL, and pSPICE) to have a safety bound, where they start dropping events/PMs when the event queuing latency is greater than or equal to 80 \% of LB, i.e., the safety bound equals to 200 milliseconds. 
We execute several runs for each experiment and show the mean value and standard deviation. 

An important factor that might influence QoR is the input event rate. Higher is the input event rate, higher is the amount of events that must be dropped and hence higher is the impact of load shedding on QoR. Additionally, other factors that might impact QoR are the query properties, e.g., the used window size. Therefore, next, we show the impact of these factors on QoR, i.e., on false negatives and positives.
Please note that applying load shedding might result in false negatives for all queries $Q_1$, $Q_2$, $Q_3$,  and $Q_4$. However, it might result in false positives only in case of $Q_3$
since $Q_3$ has a negation operator. If the negated event is dropped by the load shedder, it might result in a false positive. 

\subsubsection{Impact of Event Rate on QoR}
\label{sec:eventRate-QoR}
To evaluate the performance of \framework, we run experiments with queries $Q_1$, $Q_2$, $Q_3$, and $Q_4$.
To show the impact of input event rate, we stream both datasets to the operator with input event rates that are higher than the operator throughput $\mu$ by 20\%, 40\%, 60\%, 80\%, and 100\%  (i.e., event rate= 120\%, 140\%, 160\%, 180\%, and 200\% of the operator throughput $\mu$). Moreover, for $Q_1$, $Q_2$, and $Q_3$, we use the following window sizes, respectively: 18, 35, and 25 minutes. For $Q_4$, the used window size is 30 seconds. 
The measured operator throughput $\mu$ (without load shedding) for queries  $Q_1$, $Q_2$, $Q_3$,  and $Q_4$ are as follows: 23K, 14K, 36K, 27K events/second, respectively. 

\begin{figure*}[t]
	\centering
	\begin{subfigure}[t]{0.22\linewidth}
		\includegraphics[width=0.99\linewidth]{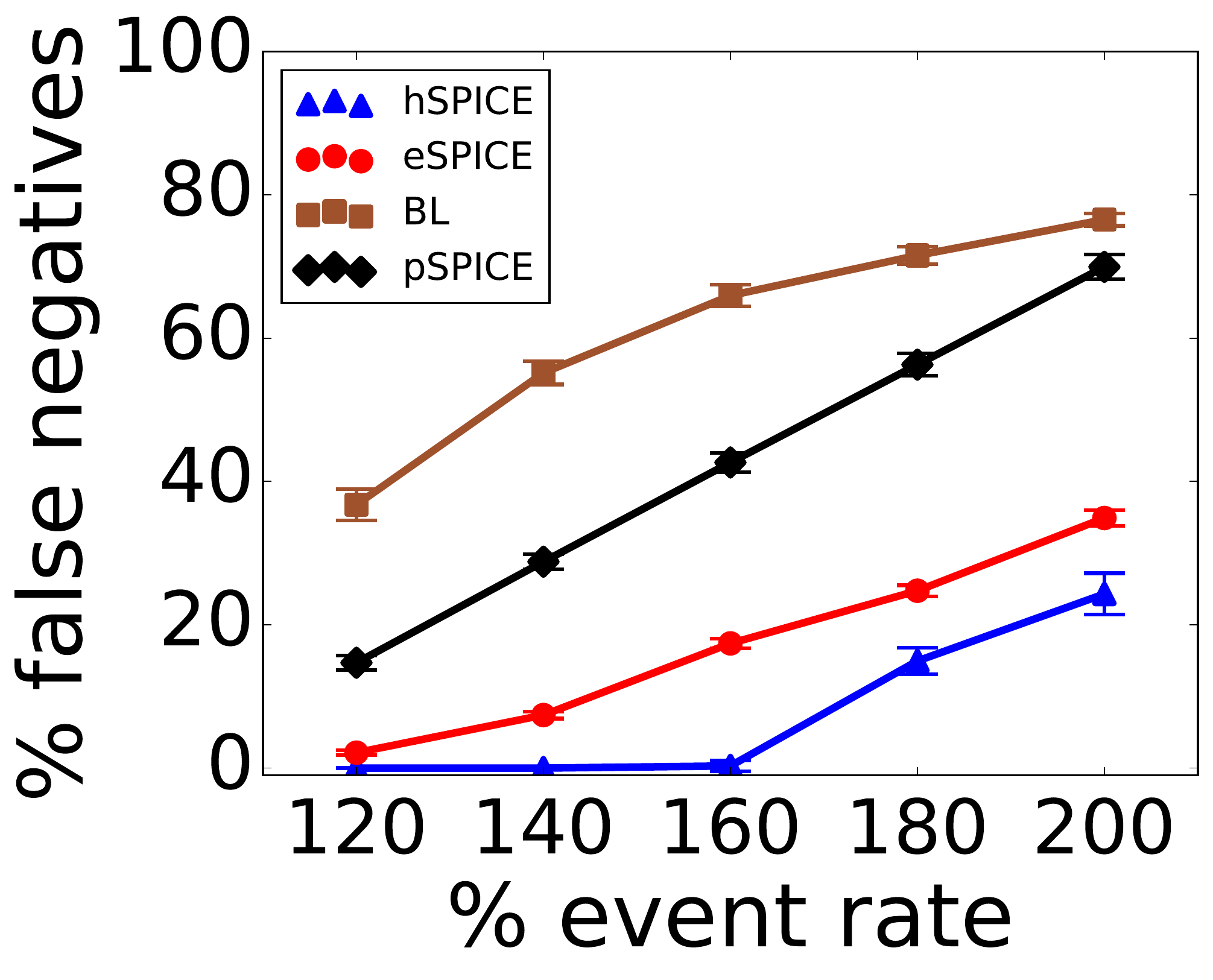} 
		\caption[]{$Q_1$}
		\label{fig:q1-fn-r}
	\end{subfigure}
	\hfil%
	\begin{subfigure}[t]{0.22\linewidth}
		\includegraphics[width=0.99\linewidth]{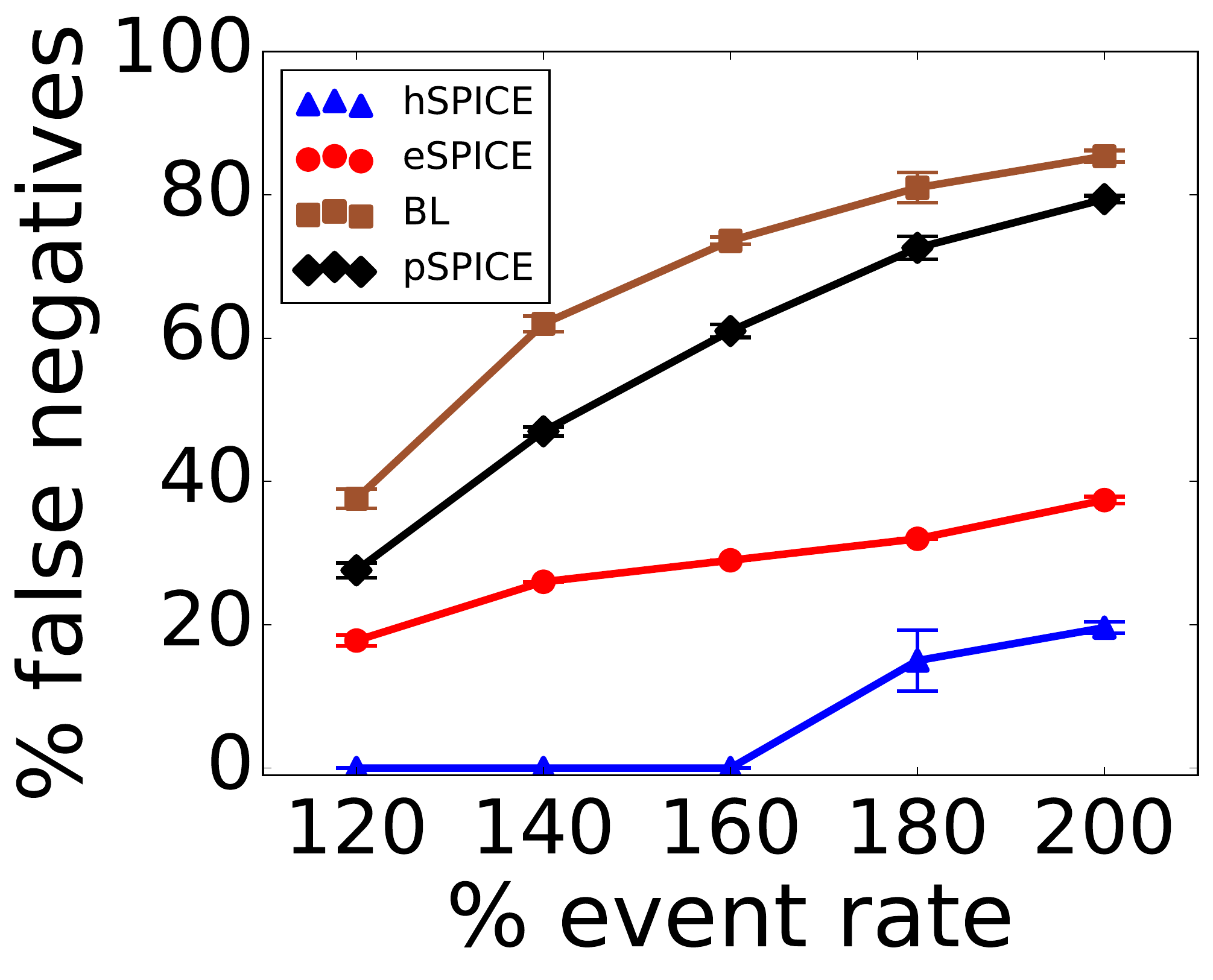}
		\caption[]{$Q_2$}
		\label{fig:q2-fn-r}
	\end{subfigure}
	\hfil%
	\begin{subfigure}[t]{0.22\linewidth}
		\includegraphics[width=0.99\linewidth]{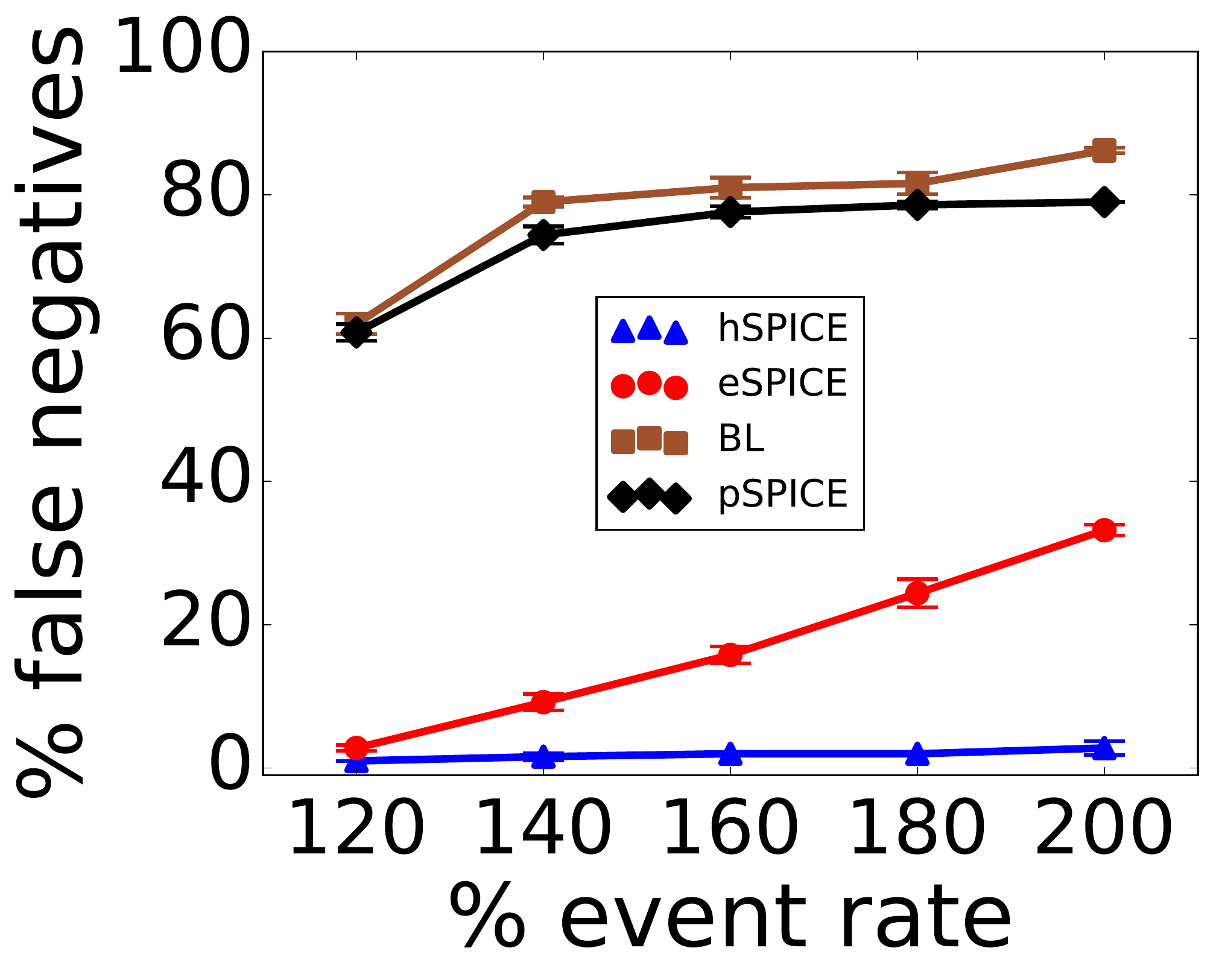}
		\caption[]{$Q_3$}
		\label{fig:q3-fn-r}
	\end{subfigure}
	\hfil%
	\begin{subfigure}[t]{0.22\linewidth}
		\includegraphics [width=0.99\linewidth]{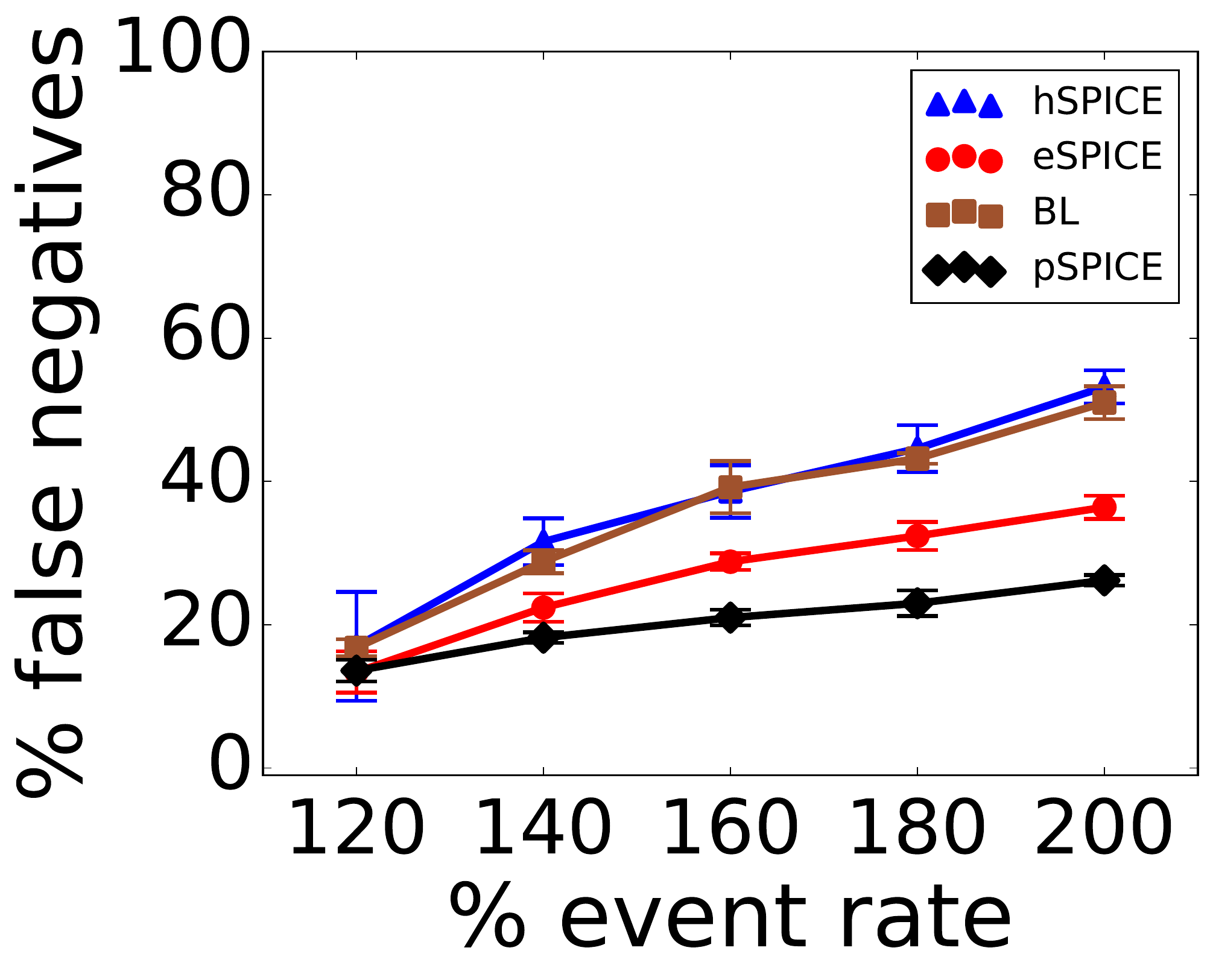}
		\caption[]{$Q_4$}
		\label{fig:q4-fn-r}
	\end{subfigure}	
	\caption{Impact of event rate on false negatives.}
	\label{fig:fn-r}
\end{figure*}

\begin{figure}[t]
	\centering
	\begin{subfigure}[t]{0.49\linewidth}
		\includegraphics[width=0.99\linewidth]{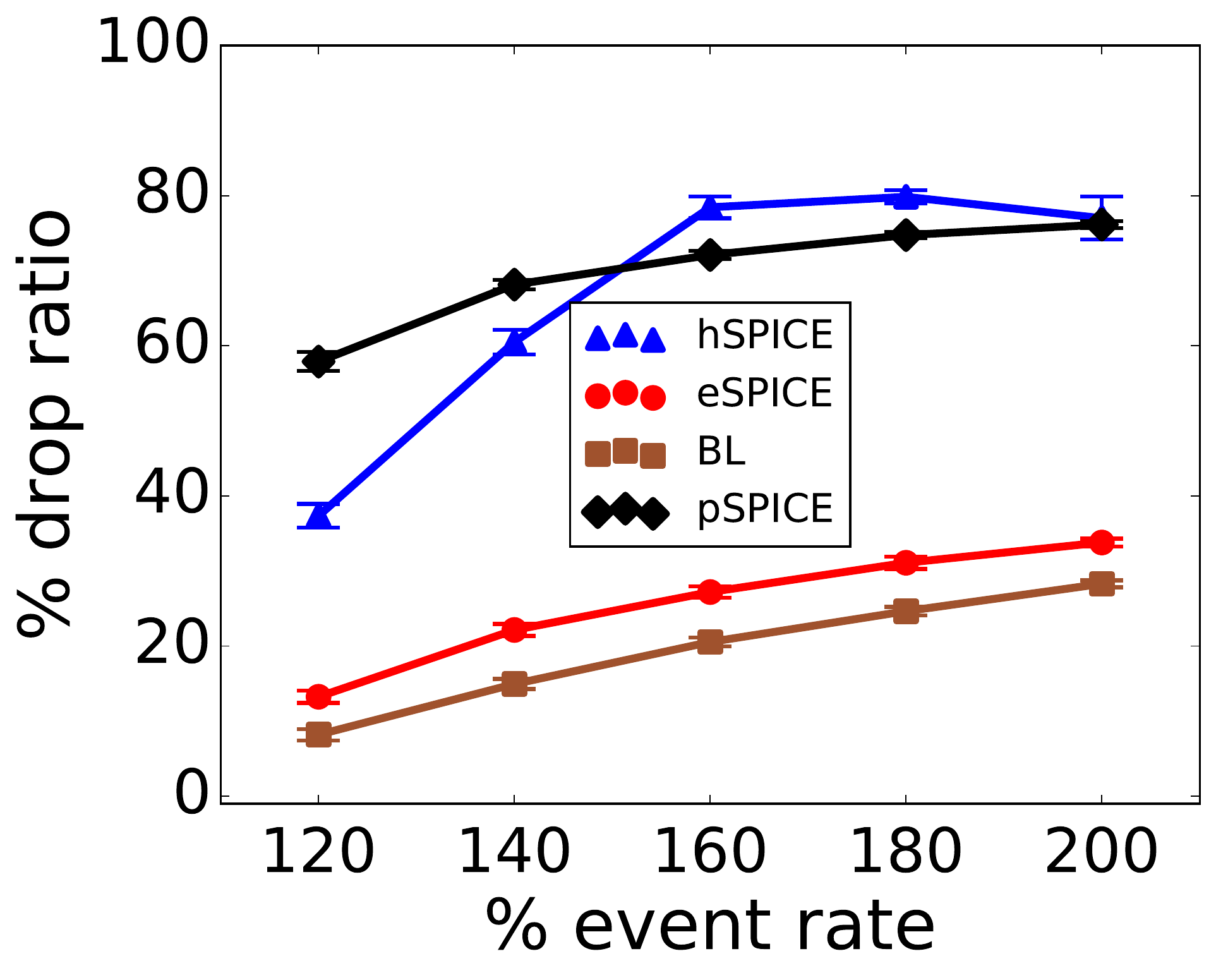}
		\caption[]{$Q_1$}
		\label{fig:q1-dr-r}
	\end{subfigure}
	\hfil%
	\begin{subfigure}[t]{0.49\linewidth}
		\includegraphics [width=0.99\linewidth]{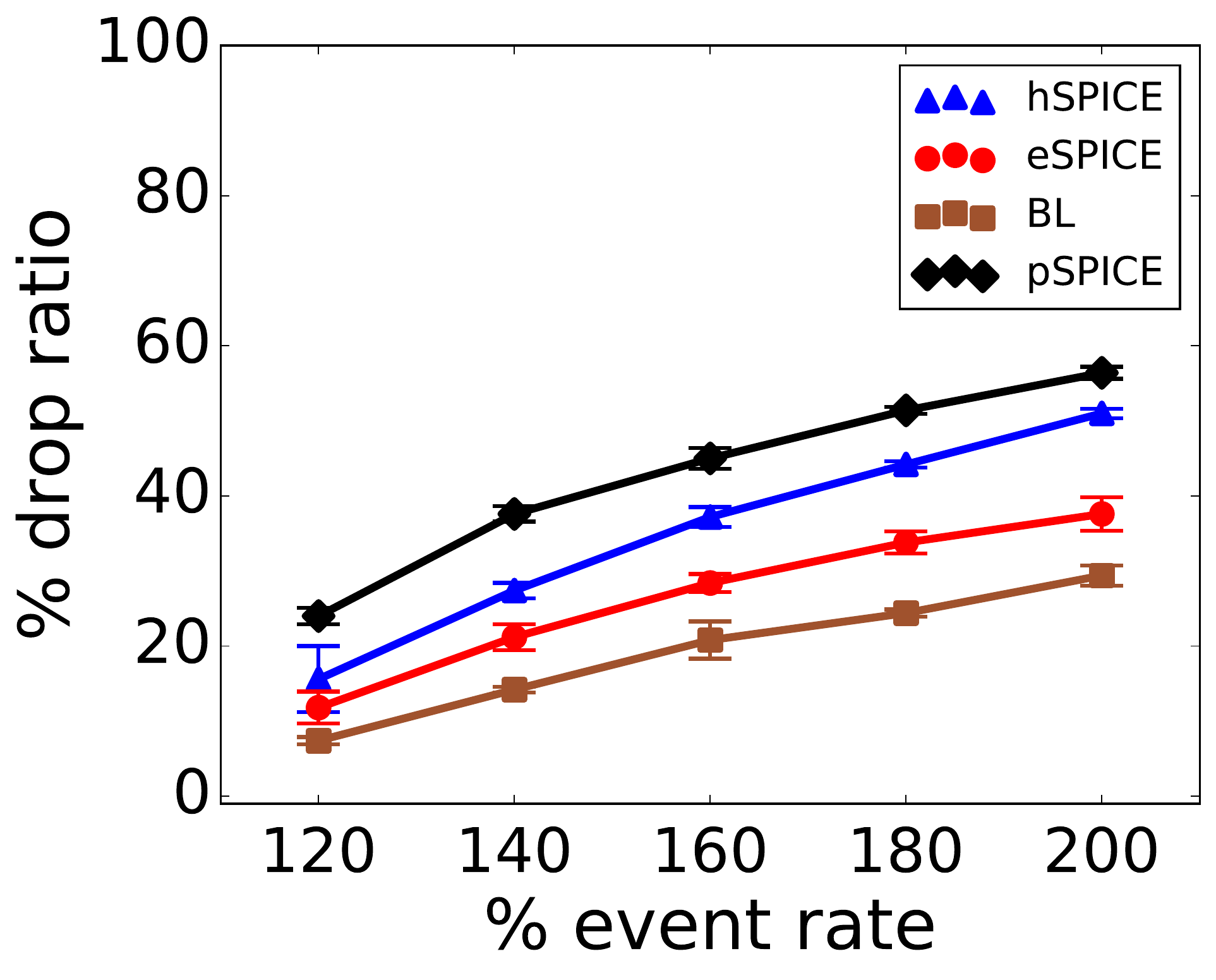}
		\caption[]{$Q_4$}
		\label{fig:q4-dr-r}
	\end{subfigure}
	
	\caption{Impact of event rate on drop ratio.}
	\label{fig:dr-r}
			\vspace{-0.4cm}
\end{figure}

\textbf{Impact on False Negatives. }
Figure \ref{fig:fn-r} depicts the impact of event rates on false negatives for all queries. Figure \ref{fig:dr-r} shows  the ratio of dropped events/PMs with different event rates for $Q_1$ and $Q_4$. We observed similar results for $Q_2$ and $Q_3$, hence we do not show them.  In both figures, the x-axis represents the event rate. The y-axis in Figure \ref{fig:fn-r} represents the percentage of false negatives while, in Figure \ref{fig:dr-r}, it represents the ratio of dropped events/PMs. 

The percentage of false negatives might increase if the input event rate increases since more events/PMs must be dropped.
Figure \ref{fig:q1-fn-r} and Figure \ref{fig:q1-dr-r} show the percentage of false negatives and the percentage of drop ratio for $Q_1$, respectively. 
As shown in Figure \ref{fig:q1-fn-r}, \framework~ has almost no impact on false negatives when the event rate is less or equal to 160\% although \framework~ drops up to 80\% of events when the event rate is 160\% as depicted in Figure \ref{fig:q1-dr-r}. Increasing the event rate by more than 160\% forces \framework~ to produce false negatives where the percentage of false negatives is 15\% and 22\% using event rates of 180\% and 200\%, respectively. The drop ratio starts to decrease when using a high event rate as shown in Figure \ref{fig:q1-dr-r} when using the event rate of 200\%. The reason behind this is that when more events should be dropped, events with high utilities might be dropped. Dropping events with high utilities might hinder opening new PMs which in turn reduces the number of events that must be dropped. Since \framework~ drops more events compared to other load shedding strategies, i.e., eSPICE and BL, the impact of shedding in \framework~ on  opening new PMs is higher which results in decreasing its drop ratio when the event rate is 200\%.
However, not opening those PMs might increase the number of false negatives. 
The percentage of false negatives caused by other load shedding strategies also increases when the event rate increases. As depicted in Figure \ref{fig:q1-fn-r}, when the event rate increases from 120\% to 200\%,  the percentage of false negatives for eSPICE, BL, and pSPICE increases from 2\% to 35\%, from 31\% to 77\%, and from 15\% to 72\%, respectively.   Moreover, the drop ratio increases with the event rate as shown in Figure \ref{fig:q1-dr-r}. This shows that \framework~ significantly outperforms all other load shedding strategies for $Q_1$ (sequence operator).
The results for $Q_2$ (sequence operator with repetition) are similar to the results for $Q_1$ as depicted in Figure \ref{fig:q2-fn-r} where \framework~ also outperforms, w.r.t. the percentage of false negatives, all other load shedding strategies.   

Figure \ref{fig:q3-fn-r} depicts the percentage of false negatives for $Q_3$ (sequence with negation operator). In $Q_3$, we limit the number of complex events to only one event per window, where the window is closed if a complex event is detected. We do that to determine the impact of the negation operator on the matching output. The performance of \framework, w.r.t. the percentage of false negatives, over $Q_3$ is  considerably better than the performance of \framework~ over $Q_1$ and $Q_2$. The reason behind this is that, in $Q_3$, there is at most one complex event per window  in comparison to $Q_1$ and $Q_2$ that detect all possible complex events in a window. Hence, in the case of $Q_3$, there exist many events in the window that have low utilities where dropping those events do not influence the percentage of false negatives. Figure \ref{fig:q3-fn-r} shows that using \framework~ with different event rates introduces almost zero false negatives. The percentage of false negatives caused by using other load shedding strategies increases with increasing event rate. In Figure \ref{fig:q3-fn-r}, the percentage of false negatives produced by eSPICE, BL, and pSPICE  increases from 3\% to 35\%, from 62\% to 84\%, and from 59\% to 79\% when increasing the event rate from 120\% to 200\%,  respectively. This shows that, for $Q_3$, \framework~ drastically reduces the percentage of false negatives compared to the other load shedding strategies.

Figures \ref{fig:q4-fn-r} and \ref{fig:q4-dr-r} show the percentage of false negatives and the percentage of drop ratio for $Q_4$ (sequence with any operator), respectively. The drop ratio in Figure \ref{fig:q4-dr-r} increases when the event rate increases. However, the drop ratio of \framework~ for $Q_4$ is lower than its drop ratio for $Q_1$. This is because the cost of processing events in $Q_4$ is higher than the cost of processing events in $Q_1$. Therefore, in $Q_4$, the overhead of performing load shedding in comparison to the event processing cost is lower which results in a low drop ratio. In Figure \ref{fig:q4-fn-r}, the percentage of false negatives caused by \framework~ increases from 13\% to 52\% when increasing the event rate from 120\% to 200\%, respectively. Whereas, the percentage of false negatives caused by eSPICE, BL, and pSPICE increases  from 13\% to 37\%, from 17\% to 50\%, and from 12\% to 26\% when increasing the event rate from 120\% to 200\%, respectively. This shows that \framework~ performs almost worse than all other load shedding strategies. The reason behind this is that the overhead of \framework~ is high in comparison to other load shedding strategies. For every event in a window, \framework~ checks whether to drop the event or not from every individual PM within the window which increases the overhead of performing load shedding in \framework. While eSPICE and BL, for example, check whether to drop the event or not from the window regardless of the number of PMs within the window which reduces the overhead of performing load shedding in these approaches. The overhead of \framework~ is high in all queries, however, the impact of \framework~ overhead is worse in $Q_4$. This is because in $Q_4$ the utility values are spread and less accurately predicted since $Q_4$ represents an \textit{any} operator in comparison to other queries that use a \textit{sequence} operator.  $Q_4$ matches an event of any type (any player) with a PM at any state, unlike the \textit{sequence} operator that matches only an event of a certain type with  a PM at a certain state. Hence, in the case of $Q_4$, the majority of events in a window have similar utilities for all PM states.

\begin{figure}[t]
	\centering
	\includegraphics [width=0.49\linewidth]{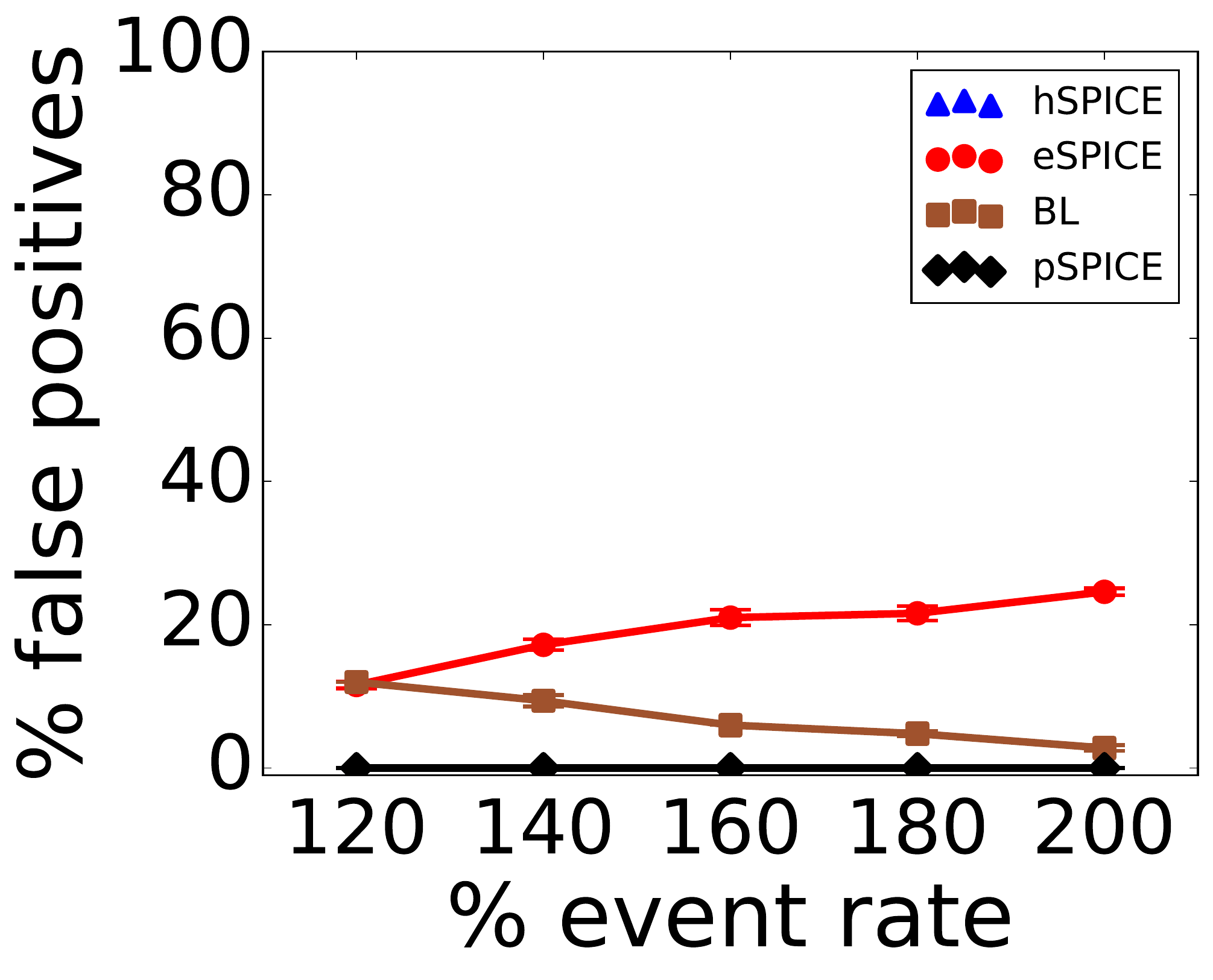}

	\caption{Impact of event rate on false positives.}
	\label{fig:q3-fp-r}
\end{figure}

\textbf{Impact on False Positives.}
As we mentioned above, only $Q_3$ (sequence with negation operator) might have false positives. Therefore, next, we analyze the impact of load shedding on the false positives using $Q_3$.  Figure \ref{fig:q3-fp-r} depicts the percentage of false positives with different event rates when using load shedding over $Q_3$. In the figure, the x-axis represents the event rate and the y-axis represents the percentage of false positives. Figure \ref{fig:q3-fp-r} shows that \framework~ performs very well with the negation operator where the percentage of false positives is almost zero for different event rates. Please recall that $Q_3$ detects at most one complex event per window.

In the figure, increasing the event rate results in increasing the percentage of false positives when using eSPICE. The percentage of false positives caused by eSPICE increases from 12\% to 24\% when increasing the event rate from 120\% to 200\%. However, the figure shows that the percentage of false positives produced when using BL  decreases from 12\% to 3\% when increasing the event rate from 120\% to 200\%. The reason behind this is that, for low event rates, BL needs to drop fewer events and hence more redundant events might exist in windows that might match the pattern. On the other hand, with a high event rate, BL must drop more events which makes it hard to have redundant events that might match the pattern.  Higher is the probability to match the pattern, higher is the probability to get false positives. 
pSPICE drops PMs, hence it can not produce any false positive.

\subsubsection{Impact of Window Size on QoR}
\begin{figure*}[t]
	\centering
	\begin{subfigure}[t]{0.22\linewidth}
		\includegraphics[width=0.99\linewidth]{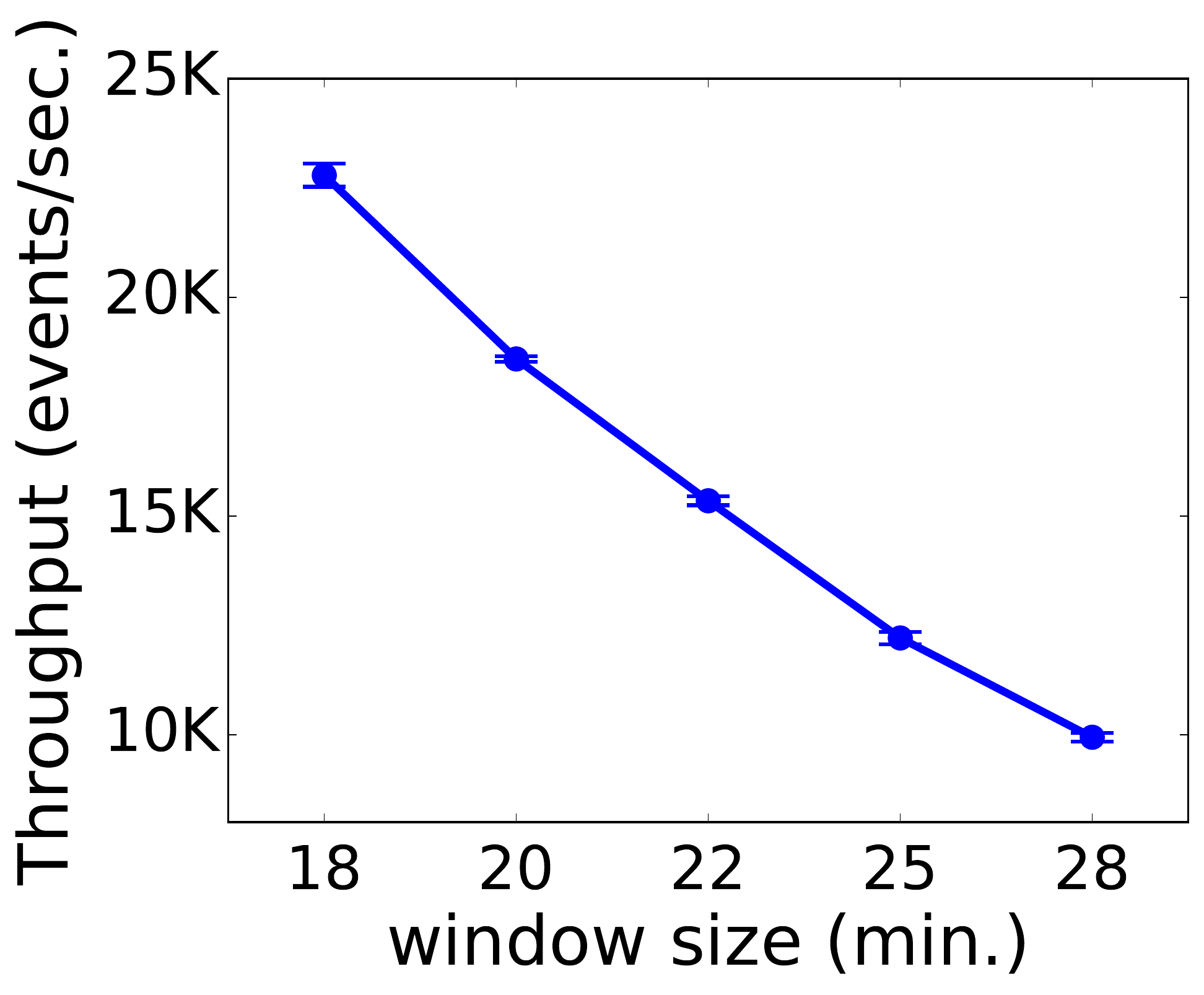}
		\caption[]{$Q_1$: operator throughput $\mu$}
		\label{fig:q1-mu-ws}
	\end{subfigure}
	\hfil%
	\begin{subfigure}[t]{0.22\linewidth}
		\includegraphics[width=0.99\linewidth]{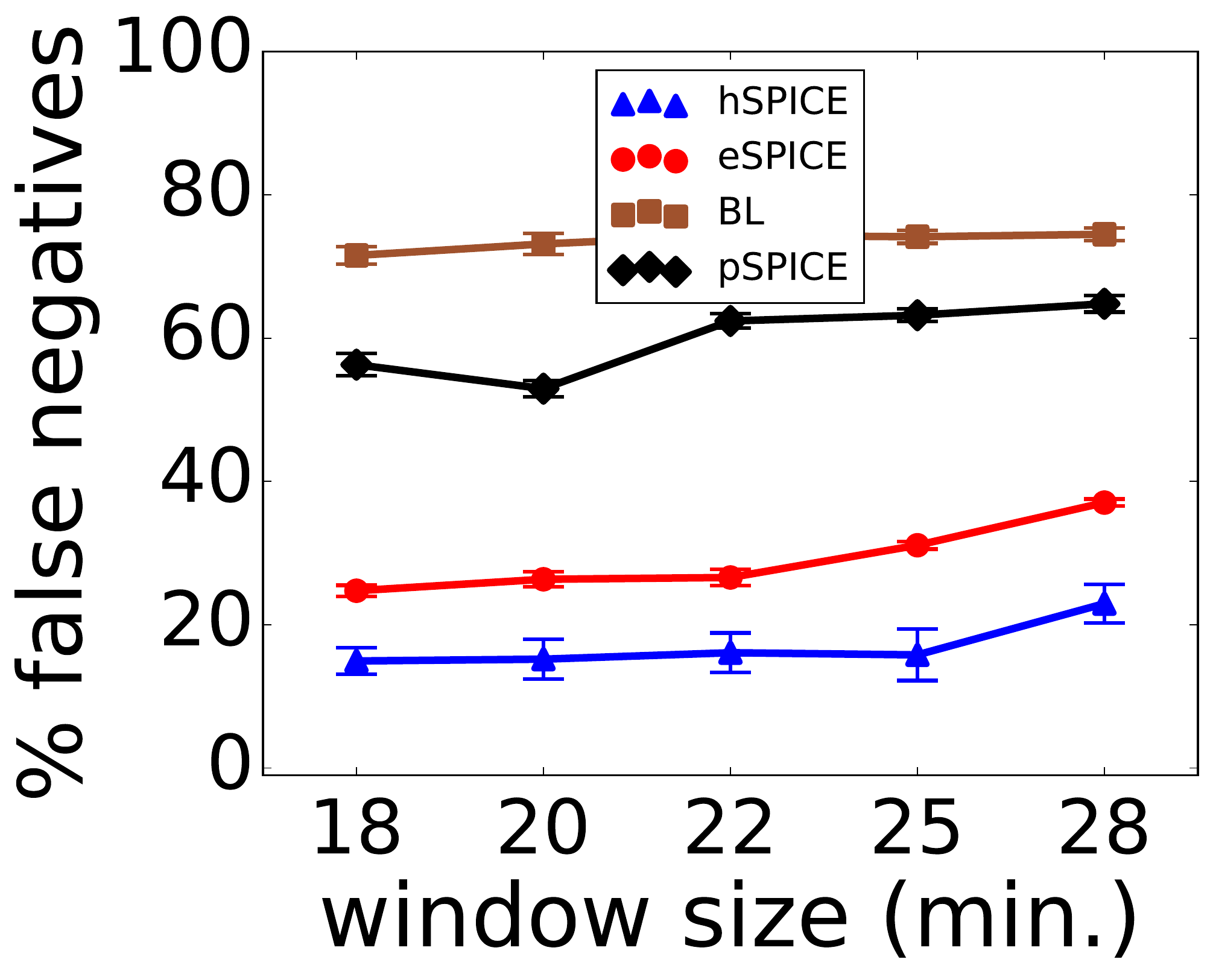}
		\caption[]{$Q_1$: false negatives}
		\label{fig:q1-fn-ws}
	\end{subfigure}
	\hfil%
	\begin{subfigure}[t]{0.22\linewidth}
		\includegraphics[width=0.99\linewidth]{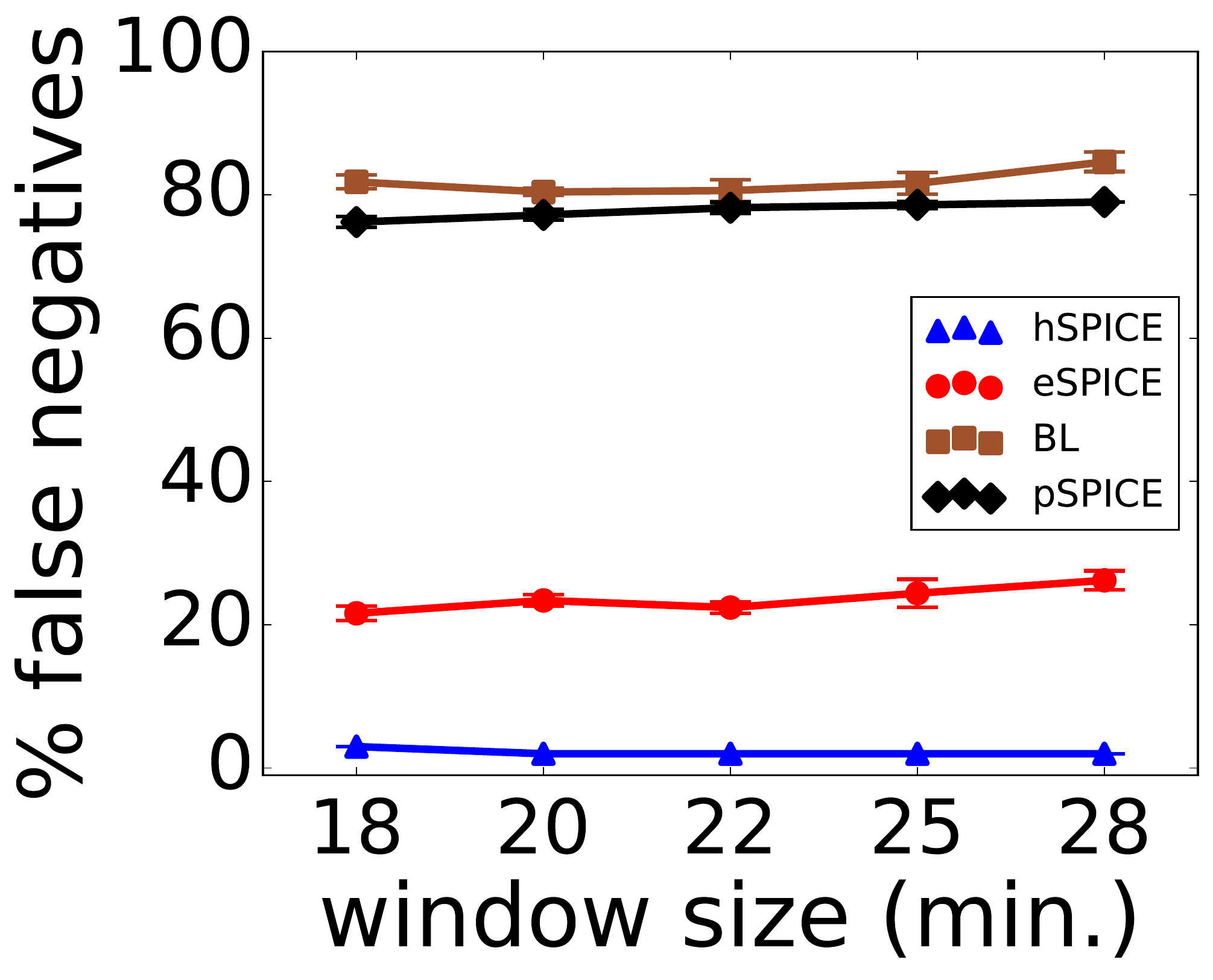}
		\caption[]{$Q_3$: false negatives}
		\label{fig:q3-fn-ws}
	\end{subfigure}
	\begin{subfigure}[t]{0.22\linewidth}
		\includegraphics[width=0.99\linewidth]{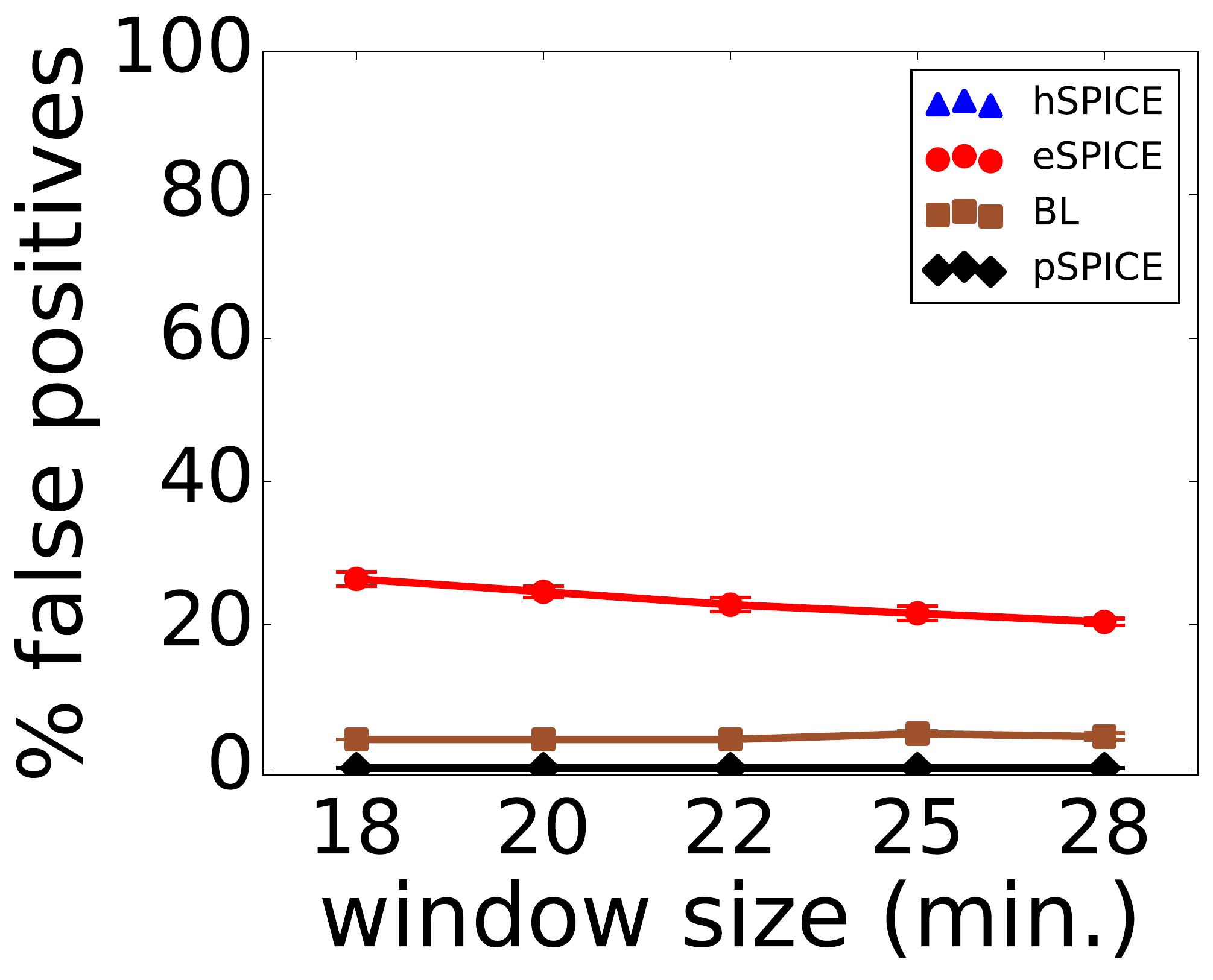}
		\caption[]{$Q_3$: false positives}
		\label{fig:q3-fp-ws}
	\end{subfigure}
	\caption{Impact of window size on QoR.}
	\label{fig:qor-ws}
\end{figure*}

In this section, we analyze the impact of window size on QoR. To do that,  we run experiments with queries $Q_1$ and $Q_3$ where we use a fixed event rate of 180\%, i.e., the input event rate is higher than the operator throughput $\mu$ by 80\%. To show the impact of window size, we vary the window size for both $Q_1$ and $Q_3$. The used window sizes for $Q_1$ and $Q_3$ are as follows: 18, 20, 22, 25, and 28 minutes. Figure \ref{fig:qor-ws} depicts the results for both queries. 
Figure \ref{fig:q1-mu-ws} shows the operator throughput $\mu$ (without load shedding) for $Q_1$ with different window sizes. If the window size increases, the number of overlapped windows increases and hence an event becomes a part of more windows. This implies that the operator throughput decreases since events must be processed in more windows.  This is observed in Figure \ref{fig:q1-mu-ws} where the operator throughput decreases from 23K to 10K when the window size increases from 18 to 28 minutes. The operator throughput for $Q_3$ has a similar behavior, hence  we do not show it. 

Figure \ref{fig:q1-fn-ws} depicts the percentage of false negatives for $Q_1$.
Increasing the window size might result in increasing the completion probability of PMs within the window. This implies that more events in the window might acquire a high utility value. Therefore,  in this case, the load shedding impact on QoR might increase. Moreover, increasing the window size might increase the number of concurrent PMs within the window where more PMs might open. This implies that the overhead of load shedding of \framework~ might increase with increasing window size since the overhead of load shedding in \framework~ is proportional to the number of PMs in windows. This might result in dropping more events hence increasing the impact on QoR. This is observed in Figure \ref{fig:q1-fn-ws} where the percentage of false negatives caused by \framework~ increases from 18\% to 21\% when the window size increases from 18 to 28 minutes. This also happens when using eSPICE where the percentage of false negatives increases from 23\% to 38\% when increasing the window size from 18 to 28 minutes. The results for pSPICE are also similar. However, the results for BL shows that the window size has almost no influence on the percentage of false negatives. This shows that \framework~ outperforms, w.r.t. the percentage of false negatives, all other load shedding strategies regardless of the used window sizes.

Figure \ref{fig:q3-fn-ws} shows the percentage of false negatives for $Q_3$. In the figure, the percentage of false negatives using \framework~ is always negligible. This is because, as we mentioned above, $Q_3$ matches at most one complex event per window, hence there might exist many events with low utilities where dropping those events has no impact on QoR. In the figure, the percentage of false negatives using eSPICE slightly increases when increasing the window size. The results for BL and pSPICE show that the percentage of false negatives stay almost stable with different window sizes. Again for $Q_3$, \framework~ outperforms, w.r.t. false negatives, all other load shedding strategies irrespective of the used window sizes. Figure \ref{fig:q3-fp-ws} depicts the percentage of false positives for  $Q_3$. The figure shows that the percentage of false positives caused by \framework~ is, again, negligible for different window sizes. In the figure, the percentage of false positives for eSPICE slightly decreases while it stays stable for BL. As mentioned above, pSPICE does not result in false positives.

\subsubsection{Maintaining Latency Bound}

\begin{figure}[t]
	\centering
	\begin{subfigure}[t]{0.49\linewidth}
		\includegraphics[width=0.99\linewidth]{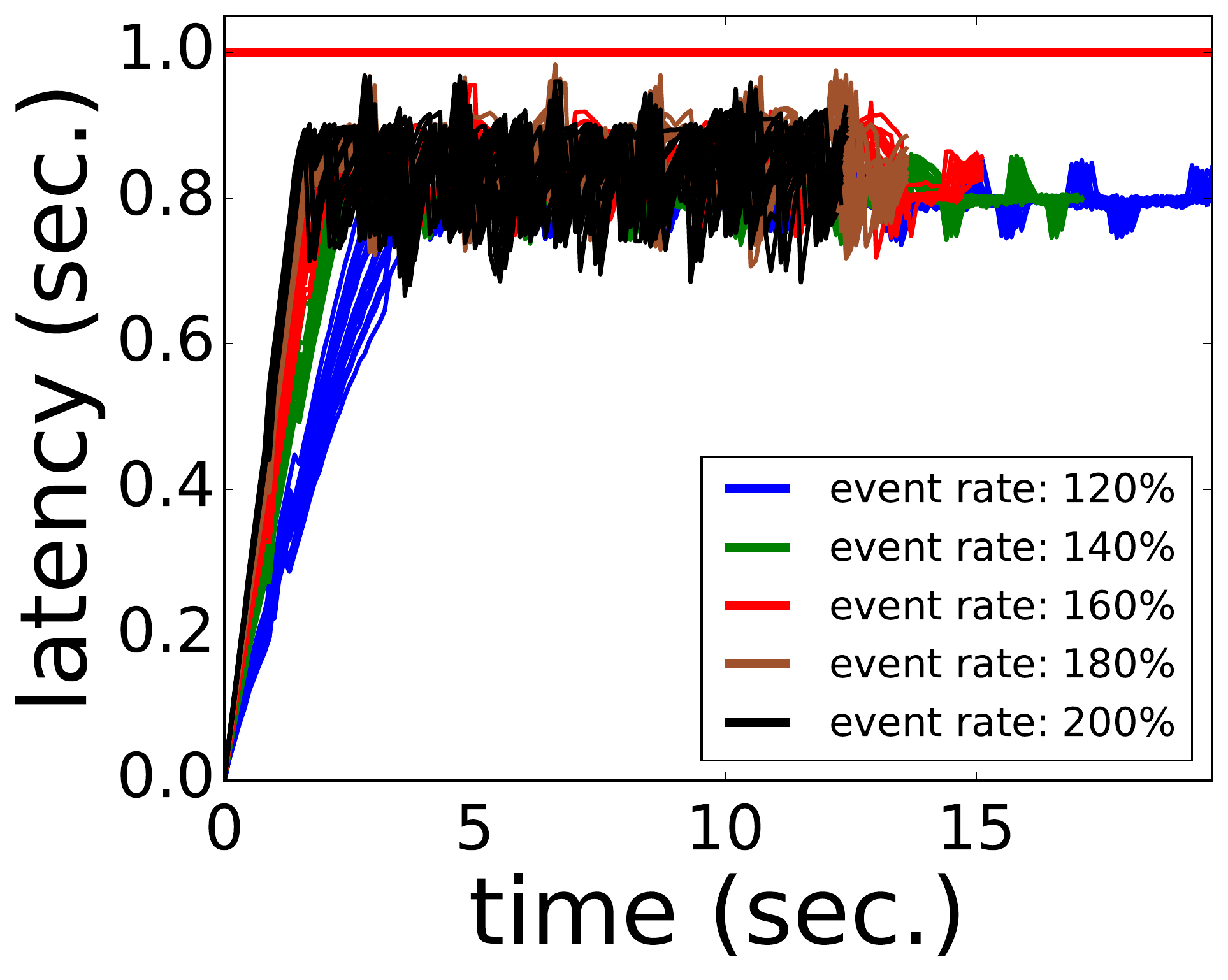}
		\vspace{-0.3cm}
		\caption[]{Q1}
		\label{fig:q1-lb}
	\end{subfigure}
	\hfil%
	\begin{subfigure}[t]{0.49\linewidth}
		\includegraphics[width=0.99\linewidth]{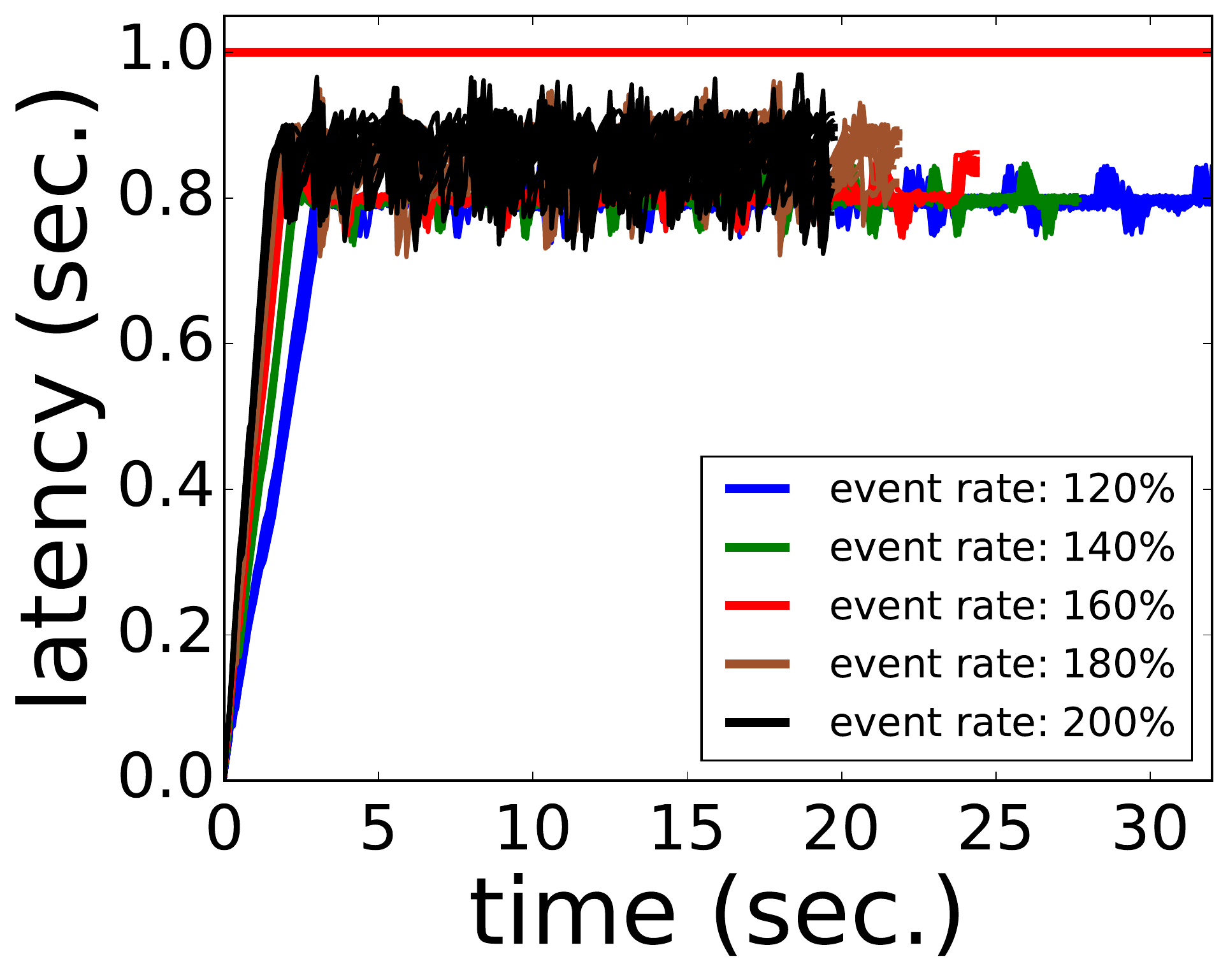}
		\caption[]{Q2}
		\label{fig:q2-lb}
	\end{subfigure}
			\vspace{-0.3cm}
	\caption{Maintaining latency bound.}
	\label{fig:lb}
			\vspace{-0.6cm}
\end{figure}

The main objective of \framework~ is to minimize the degradation in QoR while maintaining a given latency bound (LB). As mentioned above, LB is 1  second and \framework \ drops events when the event queuing latency is greater than or equal to 80\% of LB (i.e., 800 milliseconds). The event rate is an important factor that influences the ability of \framework~ to maintain LB.
Therefore, next, we show the ability of \framework~ to maintain the given latency bound (LB) with different event rates. To do that, we evaluate \framework~ with all queries using the same setting as in Section \ref{sec:eventRate-QoR}.  Figure \ref{fig:lb} shows the event latency for $Q_1$ and $Q_2$ where the event latency is the sum of the event queuing latency and the event processing latency.  In the figure, the x-axis represents the event rate and the y-axis represents the induced event latency. We observed similar results for $Q_3$ and $Q_4$, hence we do not show them.

Figures \ref{fig:q1-lb} and \ref{fig:q2-lb} depict results for $Q_1$ and $Q_2$, respectively. The figures show that \framework~ always maintains the given latency bound regardless of the event rate. In the figure, the induced event latency stays around 800 milliseconds (i.e., 80\% of LB which is used to have a safety bound).    

\subsubsection{Discussion}
\framework~ shows its ability to maintain the given latency bound while minimizing the degradation in QoR. Through extensive evaluations, we show that \framework~ outperforms, w.r.t. QoR, eSPICE, BL, and pSPICE for the majority of queries-- especially for \textit{sequence} operators. The performance of \framework~ for the \textit{any} operator is worse than the performance of other load shedding strategies.  We also show that increasing the window size  might increase the impact of \framework~ on QoR due to the following reasons. The overhead of load shedding might increase and \framework~ might have to drop more events from PMs in the window.  Moreover, the PM completion probability might increase when the window size increases, hence more events in a window might become more important/get high utilities.

\section{Related Work}
\label{sec:related_work}
Complex event processing (CEP) systems are used in many applications  to detect interesting patterns in input event streams  \cite{spectre:2017, Balkesen:2013:RRI:2488222.2488257, Zacheilas:2015, 8606633}. There exist several well-defined patterns in CEP (also called operators), e.g., sequence, negation, any,  disjunction, and conjunction \cite{4812454, Wu:2006:HCE:SASE, snoop}. In CEP systems, the input event stream is continuous and  may have a high volume. Moreover, the events are usually required to be processed in near real-time \cite{Quoc:2017:SAC:3135974.3135989, 10.1145/3361525.3361551}. Therefore, in CEP, there exist several techniques aiming to process the input events in a given latency bound such as  parallelism, optimizations, and pattern sharing \cite{spectre:2017, Balkesen:2013:RRI:2488222.2488257, Wu:2006:HCE:SASE, Ray:2016:SPS:2882903.2882947}. However, these techniques are not always sufficient or even possible, therefore, researchers propose to use load shedding.

Recently, there have been several works on load shedding in CEP \cite{espice, pspice, He2014OnLS, bo:2020}. All these approaches aim to minimize the impact of load shedding on QoR.
The approaches in \cite{espice, He2014OnLS} propose to drop events with the lowest utility from a CEP operator while the works in \cite{pspice, bo:2020} mainly drop PMs  with the lowest utility in overload situations. Besides, the authors in \cite{bo:2020} propose to drop also events if the given latency bound might be violated. In \cite{He2014OnLS}, the utility of an event depends on the event type and its frequency in the input event stream. While in \cite{espice} the utility of an event depends on  the event type and its position in the window. 
In \cite{pspice, bo:2020}, the utility of a PM depends on its completion probability and its estimated processing cost. To predict the utility of a PM, the authors propose to use  as learning features  the current state of the PM and the remaining events in the window.
Unlike all these approaches, our approach drops events from PMs where an event might have different importance for different PMs. As a result, our approach predicts the event utilities more accurately and performs dropping more precisely, thus reducing the adverse impact of load shedding  on QoR.

In the domain of approximate CEP,  the authors in \cite{Li:2016:HMF:3025111.3025121} propose a white-box approach (called RC-ACEP) to drop events from PMs in overload cases. The approach aims to minimize the degradation in QoR. They assign utilities to PMs depending on completion probabilities of the PMs-- higher is the completion probability, higher is the utility. The idea is to process input events firstly with PMs that have the highest utilities. For each newly coming input event, RC-ACEP stops processing the previous event, recalculates and sorts PM utilities, and then processes the new events with the sorted PMs. However, recalculating and sorting PM utilities for every input event imposes a high overhead. Moreover, they do not consider the importance of input events for PMs where input events might have different importance for different PMs.

Load shedding is also extensively  researched in the stream processing domain \cite{Tatbul:2006:WLS:1182635.1164196, Olston:2003:AFC:872757.872825, 3:Tatbul:2003:LSD:1315451.1315479, 3:Rivetti:2016:LSS:2933267.2933311, 3:Kalyvianaki:2016:TFF:2882903.2882943, Quoc:2017:SAC:3135974.3135989, 8622265:concept-drift, Quoc:2017:SAC:3135974.3135989, Tok:2008:SAP:1353343.1353414}.
In \cite{3:Tatbul:2003:LSD:1315451.1315479, Olston:2003:AFC:872757.872825, 8622265:concept-drift}, the authors assume that the importance of a tuple depends on the tuple's content. 
\cite{3:Tatbul:2003:LSD:1315451.1315479} assumes the mapping between the utility and tuple's content is given, for example, by an application expert, while \cite{3:Tatbul:2003:LSD:1315451.1315479, Olston:2003:AFC:872757.872825} learn this mapping online depending on the used query.    
The authors in \cite{3:Rivetti:2016:LSS:2933267.2933311} assume that the importance of a tuple depends on the processing latency of the tuple-- higher is the processing latency of a tuple, lower is its importance. Therefore, they drop those tuples that have the highest processing latencies. 
In \cite{Quoc:2017:SAC:3135974.3135989}, the authors fairly select tuples to drop from different input streams by combining two techniques: stratified sampling and reservoir sampling.
The authors in \cite{Tok:2008:SAP:1353343.1353414} also propose  to use stratified sampling and reservoir sampling to perform the approximate join.
In both these papers, the authors assume that tuples have the same utility values and impose the same processing latency. 
All these works do not capture the correlation between events in patterns which is important in CEP. For example, if the pattern is $seq(A;B)$, then events of type $A$ are only important if the stream contains events of type $B$ and vise-versa.
Our approach implicitly captures this correlation.
\vspace{-0.2cm}
\section{Conclusion}
In this paper, we proposed an efficient, lightweight load shedding strategy called \framework~ which combines the advantages of both black-box and white-box state-of-the-art load shedding strategies. In overload cases, \framework~ drops events from partial matches to maintain a given latency bound. To assign a utility value to an event for a partial match, \framework~ uses three features: 1) event type, 2) event position in the window, and 3) the current state of the partial match. By using a probabilistic model, \framework~ uses these features to predict the event utility.
Through extensive evaluations on two real-world datasets and several representative queries, we show that, for the majority of queries, \framework~ outperforms, w.r.t. QoR, state-of-the-art load shedding strategies.   Moreover, we show that \framework~ always maintains the given latency bound regardless of the incoming input event rate.

\section*{Acknowledgement}
This work was supported by the German Research Foundation (DFG) under the research grant "PRECEPT II" (BH 154/1-2 and RO 1086/19-2). The authors would like to thank Nabila Hashad for helping with implementation.

\bibliographystyle{ACM-Reference-Format}
\bibliography{paper}

\end{document}